\documentclass[11pt,letterpaper]{article}
\usepackage{amsmath}
\usepackage[T1]{fontenc}
\usepackage{lmodern}
\usepackage{microtype}
\makeatletter
\providecommand{\MT@suspend@tagging}{}
\makeatother
\usepackage[margin=1in]{geometry}
\usepackage[english]{babel}
\usepackage[ruled,linesnumbered,vlined]{algorithm2e}
\usepackage{amsmath}
\usepackage{amssymb}
\usepackage{amsthm}
\usepackage{mathtools}
\usepackage{framed}
\usepackage[numbers]{natbib}
\usepackage{tikz}
\usepackage{xcolor}
\colorlet{shadecolor}{black!6}

\usepackage[colorlinks=true, allcolors=blue]{hyperref}
\usepackage{tcolorbox}
\tcbset{
  myframe/.style={
    colback=gray!5,
    colframe=black,
    boxrule=0.8pt,
    arc=4pt,
    outer arc=4pt,
    boxsep=5pt,
    left=5pt,
    right=5pt,
    top=5pt,
    bottom=5pt,
    enhanced,
    sharp corners
  }
}

\usetikzlibrary{arrows.meta,positioning,decorations.pathreplacing}

\DeclarePairedDelimiter\floor{\lfloor}{\rfloor}
\DeclarePairedDelimiter\ceil{\lceil}{\rceil}

\newtheorem{theorem}{Theorem}[section]

\newtheorem{lemma}[theorem]{Lemma}

\theoremstyle{definition}
\newtheorem{definition}[theorem]{Definition}

\numberwithin{algocf}{section}
\numberwithin{figure}{section}
\numberwithin{equation}{section}

\newcommand{\Comp}{\mathsf{Comp}}
\newcommand{\bP}{\mathbf{P}}
\newcommand{\bX}{\mathbf{X}}
\newcommand{\bv}{\mathbf{v}}

\hypersetup{
  pdftitle={Asymptotic MMS Allocations},
  pdfauthor={}
}

\title{$(1-\varepsilon)$-MMS Allocations}
\author{
Zehan Lin\\
University of Macau\\
\texttt{yc47490@um.edu.mo}\\
\and
Shengxin Liu\\
Harbin Institute of Technology, Shenzhen\\
\texttt{sxliu@hit.edu.cn}\\
\and 
Biaoshuai Tao\\
Shanghai Jiao Tong University\\
\texttt{bstao@sjtu.edu.cn}\\
\and
Xiaowei Wu\\
University of Macau\\
\texttt{xiaoweiwu@um.edu.mo}\\
\and
Yuhao Zhang\\
Shanghai Jiao Tong University\\
\texttt{zhang\_yuhao@sjtu.edu.cn}
}

\date{}

\begin{document}
\maketitle
\begin{abstract}
We study maximin share (MMS) allocations of $m$ indivisible items among $n$ agents with additive valuations.
We prove that for both goods and chores, the optimal MMS approximation factor converges to $1$ as $n$ grows.
In particular, for every fixed $\varepsilon\in(0,1)$, sufficiently large $n$, and any $m$, we give polynomial-time algorithms that compute a $(1-\varepsilon)$-MMS allocation for goods and a $(1+\varepsilon)$-MMS allocation for chores.
Furthermore, both algorithms can be implemented using a polynomial number of bundle comparisons.
\end{abstract}

\vspace{1cm}

\paragraph{Note of Concurrent Work:} On November~23, 2026, Biaoshuai Tao informed Uriel Feige of our proof of Feige and Grinberg's asymptotic MMS conjecture for goods~\cite{feige2026redundancy}.
Feige subsequently informed us that Grinberg and he had independently proved the same conjecture.
The two groups agreed to submit their papers to arXiv on November~24, 2026, without exchanging manuscripts or proof details beforehand.
We will cite their paper and compare the results in a revised version.

\section{Introduction}\label{sec:introduction}

We study fair division, a topic that dates back to Steinhaus's work on cake cutting~\cite{steinhaus1948problem}. 
Our goal is to allocate $m$ indivisible items fairly among $n$ agents. 
In the goods setting, each agent has a nonnegative additive valuation, whereas in the chores setting, each agent has a nonnegative additive cost.
Two standard approaches formalize fairness. An envy-based notion compares agents' bundles. For example, \emph{envy-freeness} (EF)~\cite{foley1966resource} requires that no agent prefers another agent's bundle to her own.
A share-based notion compares each agent's bundle with a personal benchmark. Under \emph{proportionality} (PROP)~\cite{steinhaus1948problem}, each agent must receive at least $1/n$ of her total value for the goods or incur at most $1/n$ of her total cost for the chores. Indivisibility can rule out both guarantees, even under any fixed multiplicative relaxation. 
This motivates relaxations such as envy-freeness up to one item (EF1)~\cite{conf/sigecom/LiptonMMS04,conf/bqgt/Budish10}, envy-freeness up to any item (EFX)~\cite{journals/teco/CaragiannisKMPS19}, proportionality up to one item (PROP1)~\cite{conf/sigecom/ConitzerF017}, and the \emph{maximin share} (MMS)~\cite{conf/bqgt/Budish10}.

In this paper, we focus on MMS, a share-based notion that relaxes proportionality. Each agent imagines partitioning all items into $n$ bundles and receiving her least preferred bundle. Her MMS benchmark is the best outcome she can guarantee by choosing the partition herself. An MMS allocation gives each agent a bundle at least as good as her benchmark. Unfortunately, even this relaxation can fail with just three agents. Kurokawa, Procaccia, and Wang~\cite{KurokawaPW2018} constructed an instance for goods with no MMS allocation, and Aziz, Rauchecker, Schryen, and Walsh~\cite{conf/aaai/AzizRSW17} gave a counterexample for chores. These counterexamples motivate a multiplicative relaxation of MMS. For goods, an allocation is $\alpha$-MMS if every agent receives a bundle worth at least $\alpha$ times her MMS, where $\alpha\in[0,1]$. The best known guarantee that holds for all $n$ is $7/9$-MMS, due to Huang and Zhou~\cite{HuangZhouSevenNinths}. For chores, an allocation is $\alpha$-MMS if every agent incurs a cost of at most $\alpha$ times her MMS, where $\alpha\ge 1$. The best known guarantee that holds for all $n$ is $13/11$-MMS, due to Huang and Segal-Halevi~\cite{journals/corr/abs-2302-04581}.

How close to $1$ is the best MMS approximation factor we can guarantee for $n$ agents? For three agents, Feige, Sapir, and Tauber~\cite{conf/wine/FeigeST21} gave an instance for goods where every allocation gives some agent at most $39/40$ of her MMS and an instance for chores where every allocation costs some agent at least $44/43$ of her MMS. For four agents, Ezra and Garbuz~\cite{journals/corr/EzraG26} gave corresponding impossibility bounds of $20/21$ for goods and $31/30$ for chores. However, these constructions establish constant gaps from $1$ only for small values of $n$. For sufficiently large $n$, they showed that the best factor is at most $1-c/(\log n)^2$ for goods and at least $1+c/(\log n)^2$ for chores, for an absolute constant $c>0$. These bounds approach $1$ as $n$ grows, leaving the following question.

\begin{quote}
    \centering
    \it ``Does the optimal MMS approximation factor converge to $1$ as $n$ grows?''
\end{quote}

For goods, Feige and Grinberg~\cite{feige2026redundancy} proved that for any fixed $\delta>0$ and $\varepsilon\in(0,1)$ and sufficiently large $n$, one can select $n$ agents from a pool of at least $(1+\delta) n$ agents and give each selected agent at least $(1-\varepsilon)$ times her $n$-agent MMS.
They conjectured that for every fixed $\varepsilon\in(0,1)$ and sufficiently large $n$, a $(1-\varepsilon)$-MMS allocation exists for any number of goods.

\subsection{Our Results}

We answer this question affirmatively for goods and chores, proving Feige and Grinberg's conjecture.

\begin{tcolorbox}[colback = gray!15, colframe = gray!15]
\textbf{Main Result (Theorems~\ref{thm:cardinal-goods} and~\ref{thm:chores}).}
\emph{For every fixed $\varepsilon\in(0,1)$, all sufficiently large $n$, and any number of items, we can compute a $(1-\varepsilon)$-MMS allocation for goods and a $(1+\varepsilon)$-MMS allocation for chores in polynomial time.}
\end{tcolorbox}

We also quantify how quickly the optimal MMS approximation factors converge to one.\footnote{Note that the polynomial running-time bound depends on the fixed parameter $\varepsilon$, whereas the convergence rates below quantify the guaranteed approximation error as a function of $n$.} 
If $\alpha^*_n$ and $\rho^*_n$ denote the best MMS approximation factors guaranteed for $n$ agents over all numbers of goods and chores, respectively, our results (see Appendix~\ref{app:accuracy}) yield
\begin{equation*}
    1-\alpha^*_n=O\!\left(\left(\frac{\log\log n}{\log n}\right)^{1/8}\right)
    \qquad\text{and}\qquad
    \rho^*_n-1=O\!\left(\left(\frac{\log\log n}{\log n}\right)^{1/24}\right).
\end{equation*}

\paragraph{Comparison-Based Algorithm.}
Our algorithms for goods and chores can also be implemented using only bundle comparisons. Most existing MMS algorithms rely on numerical item values or costs~\cite{KurokawaPW2018,garg2021improved}, including those achieving $7/9$-MMS for goods~\cite{HuangZhouSevenNinths} and $13/11$-MMS for chores~\cite{journals/corr/abs-2302-04581}. 
Instead, the comparison-based query model only asks an agent which of two bundles she prefers. 
In this model, prior work achieved PROP1 together with $1/2$-MMS for goods~\cite{conf/wine/BuLLST24} and $(13/11+\varepsilon)$-MMS for chores~\cite{journals/corr/abs-2609-08687}.
Motivated by prior work, we extend our MMS guarantees to this query model.
We show that for every fixed $\varepsilon\in(0,1)$, all sufficiently large $n$, and any $m$ items, we compute a $(1-\varepsilon)$-MMS allocation for goods and a $(1+\varepsilon)$-MMS allocation for chores in polynomial time using $O(n^2\log m/\varepsilon+n^8/\varepsilon^6)$ bundle comparisons.

\medskip

For fixed $\varepsilon$ and sufficiently large $n$, this query bound matches the $\Omega(\log m)$ comparison lower bounds of computing approximate MMS allocations~\cite{conf/wine/BuLLST24}. 
More broadly, our result reveals the power of limited ordinal information, showing that polynomial bundle comparisons suffice to match the asymptotically optimal MMS guarantees achievable with full numerical valuations and costs.

\subsection{A Technical Overview}

We describe our algorithm for goods. As is standard in this line of work, we assume that the instance is IDO (meaning all agents share the same ranking of goods, with $v_i(e_1) \ge \cdots \ge v_i(e_m)$ for every agent $i$) and that each agent has a partition of the goods into $n$ bundles of value $1$. For sufficiently large $n$, our goal is to allocate an indivisible bundle of value at least $1-\varepsilon$ to every agent. Fractionally, giving every agent a $1/n$ share of each good already yields value $1$ for everyone, but this uniform fractional allocation offers little guidance on how to assign indivisible goods. To bridge this gap, we formulate a configuration LP in the spirit of Bansal and Sviridenko for max-min allocation~\cite{conf/stoc/BansalS06}. In this LP, each agent distributes one unit of fractional weight over a polynomial-size list of candidate bundles that are worth nearly $1$ to her, including the $n$ bundles in her MMS partition. Because every good in the prefix $\{e_1, \ldots, e_j\}$ is at least as valuable as $e_j$ to every agent, we can relax a bundle containing the $j$-th good into a \emph{request} for any good among the first $j$ positions. By Hall's theorem, a collection of integral requests can be matched to distinct goods whenever at most $j$ requests fall within the first $j$ positions for every $j$. Similar uses of ordered requests also appeared in Ezra and Garbuz~\cite{journals/corr/abs-2609-33234} and Feige and Grinberg~\cite{feige2026redundancy}.

With this relaxation in place, we solve an LP that minimizes a scaling factor $\lambda$ subject to the constraint that the total fractional demand in every prefix of length $j$ is at most $\lambda\cdot j$. Placing weight $1/n$ on each of an agent's $n$ partition bundles shows that $\lambda = 1$ is always feasible. A strictly smaller value of $\lambda$, say $\lambda = 0.9$, means the fractional solution leaves at least a $1-\lambda$ fraction of every prefix unused. Our analysis splits into two regimes according to the total slack $n(1-\lambda)$. 
Set $R = C\cdot \log^2 n$ for some sufficiently large constant $C$.
When $n(1-\lambda)\geq R$, we adapt the rounding argument of Feige and Grinberg~\cite{feige2026redundancy}. By keeping track of selected prefixes in each rounding step, we round the fractional solution to a single integral candidate bundle for all but $O(\log^2 n)$ agents.
Then, we match their requests to distinct goods while still leaving at least a $1-\lambda$ fraction of every prefix free. 
Since the instance is IDO, the leftover goods in these unused prefix slots are worth at least $n(1-\lambda)$ to each agent, which is more than enough to serve the remaining $O(\log^2 n)$ agents greedily.

When the slack $n(1-\lambda)$ is small, the optimal LP value is close to $1$, and LP duality provides structural leverage. Specifically, the dual LP assigns prices to goods. Using these dual prices, we identify a common prefix $M_1$ of individually valuable goods and let $M_2$ denote the remaining goods. 
For almost all agents (whom we call the \emph{ordinary} agents), every good in $M_1$ has value at least $\Omega(\varepsilon)$, while all goods in $M_2$ combined contribute only polylogarithmic value out of $n$. Since each bundle in an ordinary agent's partition has value $1$, it can contain only $O(1/\varepsilon)$ goods from $M_1$. This sparsity is crucial: it allows us to round the values of goods in $M_1$ down in small discrete steps (defined as $d_1$ in Section~\ref{sec:prefix-allocation}), while incurring only a negligible loss per bundle.

To coordinate allocations across ordinary agents, we partition $M_1$ into $O(1)$ intervals and designate a small set of positions, called \emph{checkpoints}, inside each interval. We then group together ordinary agents whose rounded values agree at all checkpoints, which allows a single representative to supply bundles for her entire group. In particular, whenever the representative requests a good lying between two checkpoints, the common IDO ordering ensures that every member of her group values the preceding checkpoint at least as highly as the representative values the requested good, up to $d_1$. Furthermore, within any given interval, most agents assign the exact same rounded value to all goods in that interval, so they are indifferent to position and can accept any good inside it.

Finally, we assemble the integral allocation from $M_1$ and $M_2$. For the ordinary agents, we select representative bundles that are already worth nearly $1$ using goods from $M_1$ alone, and we add a few bundles from a single reference partition to balance the total demand across intervals. At the same time, we cap the number of requests that require early positions within each interval, which ensures that prefix capacities are respected so that we can fill all requests with distinct goods from $M_1$ by another application of Hall's matching theorem. We then use the remaining goods in $M_2$ for two final tasks: completing the reference bundles whenever they need additional value beyond $M_1$, and serving the non-ordinary agents who place substantial value outside $M_1$.

\subsection{Related Work}\label{subsec:related}

Due to the vast literature on fair division, in the following, we focus only on the most closely related work.
For a more comprehensive review, please refer to the recent surveys~\cite{journals/sigecom/AzizLMW22,journals/ai/AmanatidisABFLMVW23}.

\paragraph{MMS Allocations for Goods.}
Since MMS allocations are not guaranteed to exist, a long line of work has steadily improved approximation guarantees for goods, starting from $2/3$-MMS~\cite{conf/sigecom/ProcacciaW14,amanatidis2017approximation,KurokawaPW2018,conf/sosa/GargMT19,journals/teco/BarmanK20} and progressing to $3/4$-MMS~\cite{ghodsi2018fair} and then to $3/4+\Omega(1/n)$~\cite{garg2021improved,akrami2023simplification}. Subsequent works broke the $3/4$ barrier with $3/4+3/3836$~\cite{Akrami_SODA24} and $10/13$~\cite{ImprovedMMS_SODA26}, and the current state of the art is a strongly polynomial-time $7/9$-MMS algorithm established by Huang and Zhou~\cite{HuangZhouSevenNinths}.
For sufficiently large $n$, Ezra and Garbuz~\cite{journals/corr/abs-2609-33234} showed that every agent can receive her MMS benchmark defined using $n+O(\sqrt{n\log n})$ bundles.
On the impossibility side, besides the small-$n$ bounds noted above, Feige, Sapir, and Tauber~\cite{conf/wine/FeigeST21} established a gap of at least $1/n^4$ from exact MMS for every $n\ge4$, which Ezra and Garbuz~\cite{journals/corr/EzraG26} strengthened to $\Omega(1/(\log n)^2)$ for large $n$.

\paragraph{MMS Allocations for Chores.}
For chores, a parallel sequence of works improved approximation guarantees from $(2-1/n)$-MMS~\cite{conf/aaai/AzizRSW17} to $4/3$-MMS~\cite{journals/teco/BarmanK20} and $11/9$-MMS~\cite{conf/sigecom/HuangL21}. The current state of the art is due to Huang and Segal-Halevi~\cite{journals/corr/abs-2302-04581}, who proved the existence of $13/11$-MMS allocations. 
For sufficiently large $n$, Lin, Wu, and Zhou~\cite{journals/corr/abs-2611-04884} obtained an ex-ante PROP lottery in which every realized allocation can be made MMS by discarding at most $o(n)$ chores.
On the negative side, in addition to the small-$n$ bounds of Feige, Sapir, and Tauber~\cite{conf/wine/FeigeST21} and Ezra and Garbuz~\cite{journals/corr/EzraG26} from the introduction, the latter proved that the optimal factor is at least $1+\Omega(1/(\log n)^2)$ for large $n$.
\paragraph{Fair Division with Limited Information.}
Several lines of work study fair division when agents' valuations cannot be elicited in full.  Under purely ordinal rankings of individual items, the best MMS approximation is of order $1/\log n$ for goods~\cite{amanatidis2016truthful,conf/ijcai/0002021}, whereas Aziz, Li, and Wu~\cite{journals/mp/AzizLW24} achieved $5/3$-MMS for chores.
For every fixed $n$, EF1 allocations for both goods and chores can be computed using $O(\log m)$ bundle comparisons~\cite{journals/corr/abs-2611-04753}.
Besides the comparison-based MMS results discussed above~\cite{conf/wine/BuLLST24,journals/corr/abs-2609-08687}, prior work has studied fair division using value queries~\cite{journals/siamdm/PlautR20,journals/siamdm/OhPS21,journals/corr/abs-2611-22376} and analyzed its communication complexity~\cite{journals/siamcomp/PlautR20,conf/sigecom/Feige25}.


\section{Preliminaries}\label{sec:preliminary}

In the following, we focus our analysis on the case of goods.
The case of chores is considered in Appendix~\ref{app:chores}.
We study the fair allocation of a set $M$ of $m$ indivisible goods among a set $N=[n]$ of $n$ agents, where $[k]=\{1,\dots,k\}$ for any positive integer $k$. We refer to any subset $S\subseteq M$ as a \emph{bundle}. Each agent $i\in N$ has a nonnegative additive valuation $v_i\colon 2^M\to\mathbb R_{\ge0}$. For a single good $e\in M$, we write $v_i(e)$ for $v_i(\{e\})$, so $v_i(S)=\sum_{e\in S}v_i(e)$ for every bundle $S\subseteq M$. An instance is denoted by $\mathcal I=(N,M,\bv)$, where $\bv=(v_1,\dots,v_n)$ is the valuation profile. For any $M'\subseteq M$ and positive integer $k$, we write $\pi_k(M')$ for the set of ordered $k$-partitions $\bP=(P_1,\dots,P_k)$ of $M'$. An \emph{allocation} is an ordered $n$-partition $\bX=(X_1,\dots,X_n)$ that assigns bundle $X_i$ to each agent $i\in N$.

\begin{definition}[MMS]\label{def:mms}
For each agent $i\in N$, her \emph{maximin share} (MMS) is defined as
\begin{equation*}
    \mu_i=\max_{\bP\in\pi_n(M)}\Bigl\{\min_{j\in[n]}\{v_i(P_j)\}\Bigr\}.
\end{equation*}
For $\alpha\in[0,1]$, an allocation $\bX$ is \emph{$\alpha$-MMS} if $v_i(X_i)\ge\alpha\cdot\mu_i$ for all $i\in N$, and \emph{MMS} if $\alpha=1$.
\end{definition}

Any partition $\bP^i\in\pi_n(M)$ attaining $\mu_i$ is called an \emph{MMS partition} of agent $i$. When $m<n$, we have $\mu_i=0$ for all $i\in N$ and every allocation is MMS. We therefore restrict our attention to instances with $m\ge n$, and throughout our analysis we assume that $n$ is sufficiently large.
Following Akrami, Garg, Sharma, and Taki~\cite{akrami2023simplification}, we call an instance \emph{normalized} if every agent $i\in N$ admits an MMS partition $\bP^i=(P_1^i,\dots,P_n^i)$ of $M$ satisfying $v_i(P_j^i)=1$ for all $j\in[n]$. In any normalized instance, every agent $i\in N$ has total value $v_i(M)=n$ and maximin share $\mu_i=1$.
An instance is \emph{identically ordered} (IDO) if the goods can be indexed as $e_1,\dots,e_m$ such that $v_i(e_1)\ge\dots\ge v_i(e_m)$ for every $i\in N$. A standard reduction gives each ordered good $e_j$ the $j$-th largest singleton value of each agent and maps allocations back to the original goods without loss in value~\cite{bouveret2016characterizing,journals/teco/BarmanK20}.

\begin{lemma}\label{lem:ido}
In polynomial time, we can reduce any goods instance $\mathcal I$ to an IDO instance $\mathcal I'$ and map any $\alpha$-MMS allocation of $\mathcal I'$ back to an $\alpha$-MMS allocation of $\mathcal I$.
\end{lemma}
Lemma~\ref{lem:ido} allows us to assume without loss of generality that the instance is IDO. For $j\in[m]$, we write $M[j]=\{e_1,\dots,e_j\}$ for the first $j$ goods in this common order.


\section{Allocation with Full Valuations}\label{sec:dichotomy}

In this section, we prove the following guarantee for goods under explicit valuations.

\begin{theorem}\label{thm:cardinal-goods}
For every fixed $\varepsilon\in(0,1)$ and all sufficiently large $n$, we can compute a $(1-\varepsilon)$-MMS allocation of goods in polynomial time.
\end{theorem}

We first normalize each agent's valuation and merge smaller items to leave $O(n)$ ordered items (Lemma~\ref{lem:reduction}). Next, we formulate a linear program (LP) that assigns fractional bundles while reserving a uniform slack among the most valuable items. When this slack is sufficiently large, we can round the LP solution to serve nearly all agents, and the leftover items provide enough value to serve the rest greedily (Lemma~\ref{lem:slack-rounding}). Otherwise, the dual reveals a common prefix of the items that holds nearly all the value for almost every agent (Lemma~\ref{lem:valuable-prefix}), and we use this structure to construct a near-MMS allocation (Theorem~\ref{thm:common-prefix-allocation}).

\subsection{Reduction and LP Framework}\label{subsec:lp-framework}

In this subsection, we first reduce any goods instance to a normalized IDO instance with $O(n)$ items (Lemma~\ref{lem:reduction}), and then formulate an LP to measure the available slack.
In addition, the reduction can be done in polynomial time with the explicit MMS partition obtained for each agent in the reduced instance (even though computing an MMS partition for a given instance is NP-hard).

\begin{lemma}\label{lem:reduction}
For fixed $0<\alpha<\beta<1$, we can reduce any goods instance $\mathcal I$ in polynomial time to a normalized IDO instance with $O(n)$ goods and explicit MMS partitions. 
Any $\beta$-MMS allocation in the reduced instance can be mapped back in polynomial time to an $\alpha$-MMS allocation of $\mathcal I$.
\end{lemma}
The proof of Lemma~\ref{lem:reduction} is deferred to Appendix~\ref{app:compression}.
%
In the following analysis, we work with a reduced normalized IDO instance with $m=O(n)$ and a fixed target $\beta\in(0,1)$, allowing implicit constants to depend on $\beta$.
We write $\bP^i=(P_1^i,\dots,P_n^i)$ for agent $i$'s MMS partition.
Furthermore, we have $v_i(M)=n$ and $v_i(e)\le1$ for all $i\in N$ and $e \in M$.
%
We fix an intermediate threshold $\sigma\in(\beta,1)$ and allow the constants in the analysis to depend on $\sigma$. For each agent $i\in N$, we construct a family $\mathcal B_i$ of \emph{candidate bundles}, where each bundle $S\in \mathcal B_i$ satisfies $v_i(S)\ge\sigma$.\footnote{Note that candidate bundles can overlap, even within the same family.}
The family $\mathcal B_i$ consists of the following three types of bundles, whose specific structure is used for the dual analysis in Section~\ref{subsec:valuable-prefix}.

\begin{itemize}
\item \emph{MMS bundles.} 
We include every bundle $B\in\bP^i$ in $\mathcal B_i$, which satisfies $v_i(B)=1>\sigma$.
\item \emph{Reduced MMS bundles.} Within each $B \in \bP^i$, we scan the items $e$ with $v_i(e)\le 1-\sigma$ in the common order and partition them into consecutive \emph{segments}. We start a new segment whenever adding the next item leads to a value above $1-\sigma$.
For each segment $B'$, we include $B\setminus B'$ in $\mathcal B_i$. Note that $v_i(B\setminus B')\ge \sigma$.
Since any two consecutive segments have total value above $1-\sigma$ and $v_i(B)=1$, each $B$ yields at most $1+\ceil{2/(1-\sigma)}$ bundles.

\item \emph{Minimal contiguous blocks.} For each $e\in M$, we include the shortest contiguous block starting from $e$ whose value to agent $i$ is at least $\sigma$, if such a block exists.
\end{itemize}

Note that the segments used to form reduced MMS bundles (the second type above) are nonempty and pairwise disjoint subsets of $M$.
Consequently, there are at most $m$ reduced MMS bundles.
Furthermore, each starting item creates at most one minimal contiguous block (the third type above). Together with the $n$ original MMS bundles, this implies $|\mathcal B_i|\le n+2m=O(n)$.
Then we solve the following LP,\footnote{Unless otherwise specified, ``the LP'' refers to this prefix relaxation in the remainder of the paper.} which assigns a variable $x_i(S)$ for each candidate bundle $S\in\mathcal B_i$ of agent $i$, denoting its assigned fraction of the bundle.
\begin{align*}
    \text{minimize}\quad & \lambda \\
    \text{subject to}\quad
    &\textstyle \sum_{S\in\mathcal B_i}x_i(S)=1 && \text{for all } i\in N, \\
    & \textstyle  \sum_{i\in N}\sum_{S\in\mathcal B_i}x_i(S)\cdot |S\cap M[j]|\le\lambda\cdot j && \text{for all } j\in[m], \\
    & x_i(S)\ge0 && \text{for all } i\in N \text{ and } S\in\mathcal B_i.
\end{align*}

The first constraint ensures that the resulting fractional allocation gives each agent a value of at least $\sigma$.
Intuitively, any item in a candidate bundle can be replaced by an earlier item without loss of value, so the second constraint caps total demand on each prefix $M[j]$ rather than on individual items. The variable $\lambda$ bounds the fraction of each prefix's capacity used, and the objective maximizes the uniform slack $1-\lambda$ across all prefixes.
This slack keeps high-value items available rather than leaving only low-value ones, ensuring enough leftover value for the subsequent greedy step.

The LP is feasible since we can set $x_i(S) = 1/n$ for each agent $i$ and each bundle $S\in \bP^i$ in the MMS partition.
This gives $\lambda=1$ and implies that the optimal solution satisfies $\lambda\le1$. Since every candidate bundle in $\mathcal B_i$ is nonempty, the prefix constraint at $j=m$ also ensures $\lambda>0$. The LP has $n$ agent constraints and $m$ prefix constraints, so we can compute in polynomial time a basic optimum $(\lambda,x)$ with at most $n+m$ positive entries in $x$.

\begin{definition}
\label{def:prefix-slack}
For an optimal solution $(\lambda,x)$ of the LP, its \emph{total slack} is defined as $\Delta=n\cdot(1-\lambda)$.
\end{definition}

We set $L=\ceil{\log_2 n}$ and use $R=768L^2=O(\log^2 n)$ as the rounding loss budget. We split our analysis into two regimes. When $\Delta\ge2R$, Section~\ref{subsec:slack-rounding} rounds $x$ and allocates the leftover items to guarantee every agent value at least $\sigma$. When $\Delta<2R$, Section~\ref{subsec:valuable-prefix} uses the dual LP to show that almost all agents agree on a common valuable prefix of items, and Sections~\ref{sec:prefix-allocation} and~\ref{subsec:completing-allocation} utilize this property to compute the near-MMS allocation.

\subsection{Large Slack: \texorpdfstring{$\Delta\ge 2R$}{Delta >= 2R}}\label{subsec:slack-rounding}
In this subsection, we handle the case $\Delta\ge2R$, where the LP leaves substantial unused capacity across every prefix. Intuitively, this uniform slack allows us to satisfy almost all agents while setting aside a balanced pool of both large and small items. We adapt the iterative rounding method of Feige and Grinberg~\cite{feige2026redundancy} to leave only $O(\log^2 n)$ agents unserved. The leftover pool then provides enough value to serve them greedily (see Algorithm~\ref{alg:slack} and Lemma~\ref{lem:slack-rounding}).


\paragraph{Algorithm Overview.}
Algorithm~\ref{alg:slack} constructs the allocation in three steps. First, we use the iterative checkpoint rounding~\cite{feige2026redundancy} to round $x$ into an integral bundle selection that preserves every prefix constraint while leaving at most $R$ agents unserved. Second, for these selected bundles, we replace each requested item with a distinct item at or before its position in the common order, guaranteeing value at least $\sigma$ to every served agent. 
Since at least a $1-\lambda$ fraction of each prefix remains unassigned, the leftover pool has value at least $\Delta$ to every agent. 
Finally, we serve the remaining agents greedily from this pool by a simple bag-filling. 
Each bundle stops as soon as it reaches value $\sigma$ for some unserved agent, so $\Delta\ge2R$ suffices to satisfy them all.

\medskip

\begin{algorithm}[!t]
\caption{Checkpoint Rounding and Bundle Filling with Total Slack}
\label{alg:slack}
\linespread{1.5}
\KwIn{$\mathcal{I} = (N,M,\bv)$, candidate bundles $\{\mathcal B_i\}_{i\in N}$, target $\sigma$, and a basic optimum $(\lambda,x)$}
\KwOut{Allocation $\bX=(X_1,\dots,X_n)$}
\tcp{Step 1: Iterative checkpoint rounding}
Set $\mathcal F\gets\{(i,S):0<x_i(S)<1\}$\;
\While{$\mathcal F\ne\emptyset$}{
    Set $\omega\gets 16(1+\ceil{\log_2(m\cdot|\mathcal F|)})/|\mathcal F|$\;
    \If{$\omega\ge 1/2$}{
        \textbf{break}\;
    }
    Form $\tilde x$ from $x$ by setting
    $\tilde x_i(S)\gets\max\{0,x_i(S)-\omega\}$ for all $(i,S)\in\mathcal F$\;
    $x\gets$ an optimal vertex of $\mathrm{LP}(\tilde x)$, and update $\mathcal F$\;
}
Set $x_i(S)\gets 0$ for all $(i,S)\in\mathcal F$\;
\tcp{Step 2: Deadline assignment}
Sort all requests $(i,e)$ with $x_i(S)=1$ and $e\in S$ by item position\;
For the $k$-th request $(i,e)$, assign the $\ceil{k/\lambda}$-th item of $M$ to $X_i$\;
$U\gets M\setminus\bigcup_{i\in N}X_i$\;
\tcp{Step 3: Greedy bundle filling}
$N^-\gets\bigl\{i\in N:\sum_{S\in\mathcal B_i}x_i(S)=0\bigr\}$\;
\While{$N^-\ne\emptyset$}{
    Let $B'\subseteq U$ be the shortest prefix of $U$ with $v_i(B')\ge\sigma$ for some $i\in N^-$\;
    $X_i\gets B'$, $U\gets U\setminus B'$, $N^-\gets N^-\setminus\{i\}$\;
}
Add $U$ to $X_1$;\\
\Return $\bX=(X_1,\dots,X_n)$\;
\end{algorithm}

Concretely, Step~1 eliminates all fractional entries of $x$ while preserving every prefix bound by repeatedly updating $x$ to an optimal vertex of an auxiliary \emph{rounding LP}.
In a basic feasible solution of a linear program, once all integral entries are fixed, the number of fractional entries is at most the number of tight constraints involving them.
If a rounding LP includes all $m$ prefix constraints, this upper bound is $O(m)$ and does not guarantee that the fractional support $\mathcal F=\{(i,S):0<x_i(S)<1\}$ shrinks.
Instead, each rounding LP includes prefix constraints only at a small set of positions in $[m]$, called \emph{checkpoints} in $\mathcal C(\tilde x)$.
For each checkpoint, the rounding LP requires the prefix load not to increase.
However, because other positions are unconstrained in the LP, the load on a non-checkpoint prefix $M[j]$ can still increase.
To preserve the prefix bounds at these positions, Step~1 slightly reduces every fractional entry before solving the LP, creating a small slack below $\lambda\cdot j$ on each $M[j]$ to offset any such increase.

Specifically, in each iteration with nonempty $\mathcal F$, we set $\omega=16\bigl(1+\ceil{\log_2(m\cdot|\mathcal F|)}\bigr)/|\mathcal F|$.
When $\omega<1/2$, we keep all integral entries fixed and reduce each fractional entry by $\omega$:
\begin{equation*}
    \tilde x_i(S)=\max\{0,x_i(S)-\omega\}\qquad \text{for all } (i,S)\in\mathcal F.
\end{equation*}

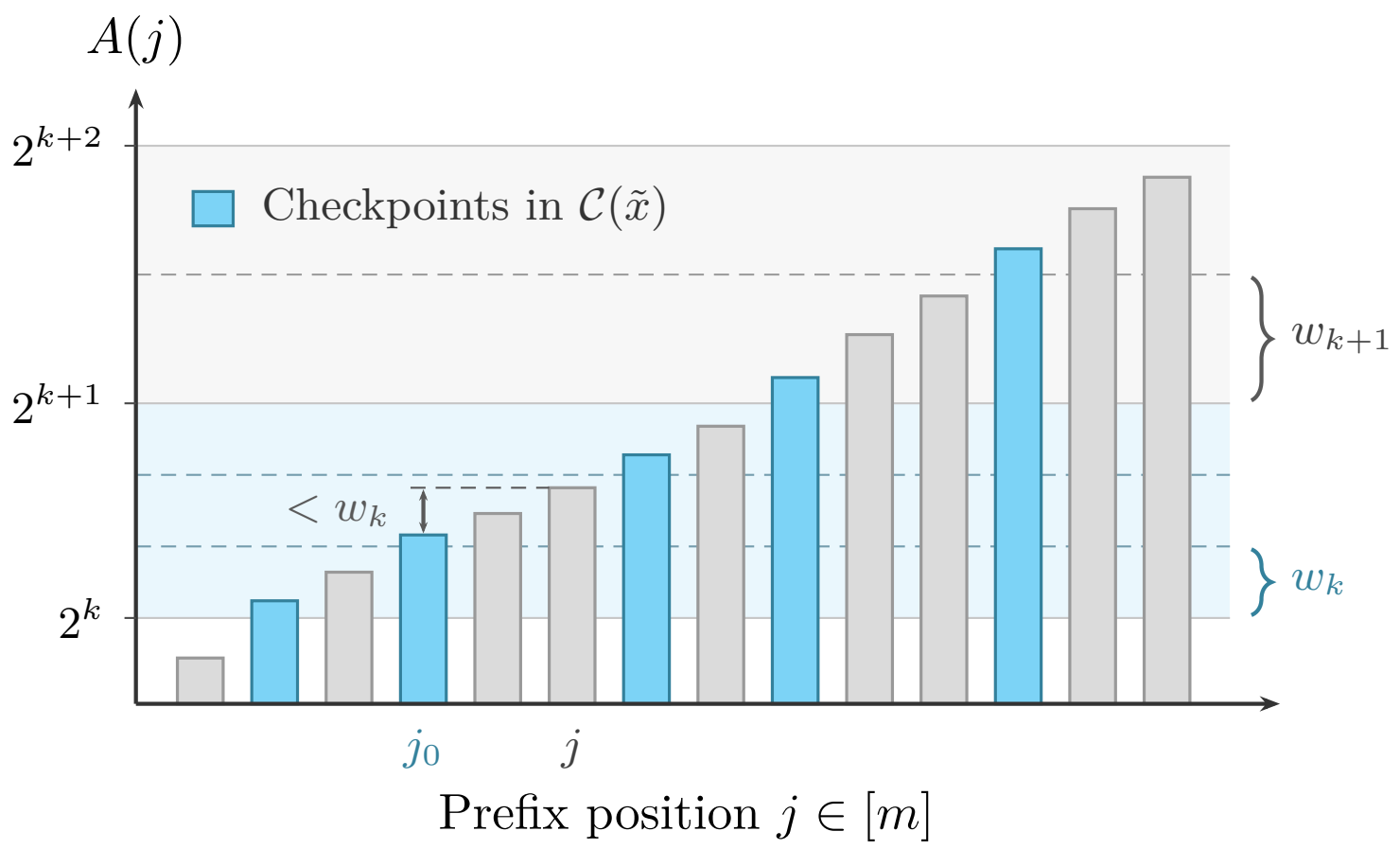
\begin{figure}[htbp]
  \centering
  \begin{tikzpicture}[
    x=1cm,y=1cm,font=\footnotesize,
    every node/.style={inner sep=1.5pt}
  ]
    \fill[cyan!7] (0,0.6) rectangle (7.65,2.1);
    \fill[black!3] (0,2.1) rectangle (7.65,3.9);

    \draw[black!22,thin] (0,0.6) -- (7.65,0.6);
    \draw[black!22,thin] (0,2.1) -- (7.65,2.1);
    \draw[black!22,thin] (0,3.9) -- (7.65,3.9);

    \draw[cyan!45!black,densely dashed,opacity=0.7] (0,1.1) -- (7.65,1.1);
    \draw[cyan!45!black,densely dashed,opacity=0.7] (0,1.6) -- (7.65,1.6);
    \draw[black!38,densely dashed] (0,3.0) -- (7.65,3.0);

    \foreach \x/\h/\fillc/\drawc in {
      0.45/0.32/black!14/black!40,
      0.97/0.72/cyan!45!white/cyan!55!black,
      1.49/0.92/black!14/black!40,
      2.01/1.18/cyan!45!white/cyan!55!black,
      2.53/1.33/black!14/black!40,
      3.05/1.51/black!14/black!40,
      3.57/1.74/cyan!45!white/cyan!55!black,
      4.09/1.94/black!14/black!40,
      4.61/2.28/cyan!45!white/cyan!55!black,
      5.13/2.58/black!14/black!40,
      5.65/2.85/black!14/black!40,
      6.17/3.18/cyan!45!white/cyan!55!black,
      6.69/3.46/black!14/black!40,
      7.21/3.68/black!14/black!40
    }{
      \fill[\fillc] (\x-0.16,0) rectangle (\x+0.16,\h);
      \draw[\drawc,semithick] (\x-0.16,0) rectangle (\x+0.16,\h);
    }

    \draw[-{Stealth[length=4pt]},thick,black!80] (0,0) -- (8.0,0);
    \draw[-{Stealth[length=4pt]},thick,black!80] (0,0) -- (0,4.3);
    \node[anchor=south] at (0,4.40) {\small $A(j)$};
    \node[anchor=north] at (3.85,-0.55) {\small Prefix position $j\in[m]$};

    \foreach \y/\lab in {0.6/{$2^k$}, 2.1/{$2^{k+1}$}, 3.9/{$2^{k+2}$}}{
      \draw[black!70] (-0.08,\y) -- (0,\y);
      \node[anchor=east] at (-0.16,\y) {\lab};
    }

    \node[anchor=north,text=cyan!55!black] at (2.01,-0.12) {$j_0$};
    \node[anchor=north,text=black!75] at (3.05,-0.12) {$j$};

    \draw[black!65,densely dashed] (1.92,1.51) -- (2.89,1.51);
    \draw[{Stealth[length=2.3pt]}-{Stealth[length=2.3pt]},black!65,thick]
      (2.01,1.19) -- (2.01,1.50);
    \node[anchor=east,text=black!65] at (1.84,1.35) {$<w_k$};

    \draw[decorate,decoration={brace,amplitude=4pt,mirror},cyan!55!black,thick]
      (7.80,0.62) -- (7.80,1.08)
      node[midway,right=6pt,text=cyan!55!black] {$w_k$};
    \draw[decorate,decoration={brace,amplitude=4pt,mirror},black!65,thick]
      (7.80,2.12) -- (7.80,2.98)
      node[midway,right=6pt,text=black!75] {$w_{k+1}$};

    \fill[cyan!45!white] (0.40,3.34) rectangle (0.68,3.58);
    \draw[cyan!55!black,semithick] (0.40,3.34) rectangle (0.68,3.58);
    \node[anchor=west,text=black!80] at (0.82,3.46) {Checkpoints in $\mathcal C(\tilde x)$};
  \end{tikzpicture}
  \caption{Checkpoint placement for $\tilde x$. The first bar reaching each threshold is selected into $\mathcal C(\tilde x)$, so any position $j$ exceeds its preceding checkpoint $j_0\le j$ by $A(j)-A(j_0)<w_k$.} 
  \label{fig:checkpoints}
\end{figure}

Below, we define the rounding LP, denoted by $\mathrm{LP}(\tilde x)$.
This program keeps all integral entries fixed, and enforces prefix bounds only at a subset $\mathcal C(\tilde x)$ of $[m]$ to be formally defined later.
\begin{align*}
\text{maximize}\quad & \textstyle \sum_{i\in N}\sum_{S\in\mathcal B_i}x'_i(S) \\
    \text{subject to}\quad
    & \textstyle \sum_{S\in\mathcal B_i}x'_i(S)\le 1 && \text{for all } i\in N, \\
    & x'_i(S)=\tilde x_i(S) && \text{for all } (i,S) \text{ with } \tilde x_i(S)\in\{0,1\}, \\
    & \textstyle \sum_{i\in N}\sum_{S\in\mathcal B_i}\bigl(x'_i(S)-\tilde x_i(S)\bigr)\cdot |S\cap M[j]|\le 0 && \text{for all } j\in\mathcal C(\tilde x), \\
    & x'_i(S)\ge0 && \text{for all } (i,S).
\end{align*}
For each $j\in[m]$, the entries remaining fractional in $\tilde x$ have total prefix count
\begin{equation*}
    A(j)=\sum_{(i,S):\,0<\tilde x_i(S)<1}|S\cap M[j]|.
\end{equation*}
Because each of these entries decreases by $\omega$, the transition from $x$ to $\tilde x$ creates a slack of at least $\omega\cdot A(j)$ below $\lambda\cdot j$ on $M[j]$.
If a checkpoint $j_0\le j$ prevents the load on $M[j_0]$ from increasing in the next LP step, then the load on $M[j]$ can increase by at most $A(j)-A(j_0)$ from $\tilde x$, so larger values of $A(j)$ tolerate wider gaps between checkpoints without exceeding $\lambda\cdot j$.
We therefore group all possible positive values $\{1,\dots,m\cdot|\mathcal F|\}$ of $A(j)$ into $O(\log(m\cdot|\mathcal F|))$ ranges $[2^k,2^{k+1})$ for integers $k\ge0$ with $2^k\le m\cdot|\mathcal F|$, and set the threshold spacing to $w_k=\max\{1,\floor{\omega\cdot 2^k}\}$.
Within each range $[2^k,2^{k+1})$, we place thresholds at $2^k,2^k+w_k, 2^k + 2w_k, \dots$ below $2^{k+1}$, which uses $O(1/\omega)$ thresholds per range and $O\bigl(\log(m\cdot|\mathcal F|) / \omega \bigr)$ thresholds overall.
For each threshold, we include in the checkpoint set $\mathcal C(\tilde x)$ the smallest position $j\in[m]$ where $A(j)$ is at least that threshold, if one exists (see Figure~\ref{fig:checkpoints} for illustration).
This completes the definition of the rounding LP.
%
By our choice of $\omega$, the set $\mathcal C(\tilde x)$ has size at most $|\mathcal F|/4$, which reduces $|\mathcal F|$ by a constant fraction in each iteration while decreasing the sum of entries in $x$ by at most $\omega\cdot|\mathcal F|=O(\log n)$.

In Step~2, every variable lies in $\{0,1\}$, and each item $e$ in a candidate bundle with coefficient $1$ creates a request whose deadline is its position in the common order. Under the common order, any item at or before that position satisfies the request without decreasing the agent's value. We sort all requests by deadline and assign the $\ceil{k/\lambda}$-th item of $M$ to the $k$-th request. Since $\lambda\le1$, these positions are strictly increasing and hence distinct. Moreover, if the $k$-th request has deadline $j$, the preserved prefix constraint gives $k\le\lambda\cdot j$ and therefore $\ceil{k/\lambda}\le j\le m$, so the assigned item always exists and satisfies the request. 
Since the assigned positions are $\ceil{k/\lambda}$, each prefix $M[j]$ contains at most $\floor{\lambda\cdot j}$ assigned items, leaving at least $(1-\lambda)\cdot j$ unassigned items in $M[j]$ for Step~3.

In Step~3, the set $N^-$ collects the agents left unserved by Step~1.
While $N^-$ is nonempty, we assign the shortest prefix $B'\subseteq U$ satisfying $\max_{i\in N^-}\{v_i(B')\}\ge\sigma$ to a qualifying agent $i\in N^-$ and remove $B'$ from $U$ and $i$ from $N^-$.
Finally, any leftover items in $U$ are added to $X_1$.


\begin{lemma}\label{lem:prefix-load}
Throughout Step~1 of Algorithm~\ref{alg:slack}, for any $j \in [m]$, the prefix $M[j]$ satisfies
\begin{equation*}
    \sum_{i\in N}\sum_{S\in\mathcal B_i}x_i(S)\cdot|S\cap M[j]|\le\lambda\cdot j.
\end{equation*}
\end{lemma}
\begin{proof}
We proceed by induction on the iterations of Step~1. The bound holds initially because $(\lambda,x)$ is feasible for the LP. In an iteration with $\omega\ge1/2$, setting every entry in $\mathcal F$ to $0$ cannot increase any prefix load, which preserves the bound.
When $\omega<1/2$, for each $j\in[m]$, the entries remaining fractional in $\tilde x$ have prefix count $A(j)$.
Because each of these entries decreases by $\omega$ while all other entries do not increase, the load of $\tilde x$ on $M[j]$ is at most $\lambda\cdot j-\omega\cdot A(j)$. Moreover, $\tilde x$ is feasible for $\mathrm{LP}(\tilde x)$, and in its basic optimum $x'$ only entries with $0<\tilde x_i(S)<1$ can change, each by less than $1$.
For any $j\in[m]$, if $A(j)=0$ then the load on $M[j]$ is unchanged from $\tilde x$ to $x'$. Otherwise, $A(j)\in[2^k,2^{k+1})$ for a unique integer $k\ge0$ with $2^k\le m\cdot|\mathcal F|$, and the largest threshold at most $A(j)$ yields a checkpoint $j_0\in\mathcal C(\tilde x)$ with $j_0\le j$ and $A(j)-A(j_0)<w_k$. Because the load on $M[j_0]$ cannot increase from $\tilde x$ to $x'$, the load on $M[j]$ increases by at most $A(j)-A(j_0)$. When $w_k=1$, this nonnegative integer difference is $0$. If $w_k\ge2$, then we have
\begin{equation*}
    A(j)-A(j_0)<w_k=\floor{\omega\cdot 2^k}\le\omega\cdot 2^k\le\omega\cdot A(j).
\end{equation*}
In both cases, this increase is at most $\omega\cdot A(j)$, so the load of $x'$ on $M[j]$ stays at most $\lambda\cdot j$.
\end{proof}

\begin{lemma}\label{lem:step1-rounding}
Step~1 terminates in polynomial time with an integral weight vector $x$ and $|N^-|\le R$.
\end{lemma}
\begin{proof}
In any iteration with $\omega<1/2$, there are at most $1+\ceil{\log_2(m\cdot|\mathcal F|)}$ integers $k\ge0$ with $2^k\le m\cdot|\mathcal F|$. For each such $k$, the number of thresholds is $2^k<1/\omega$ if $\omega\cdot 2^k<1$, and at most $1+2^k/\floor{\omega\cdot 2^k}\le 1+2/\omega<4/\omega$ otherwise (since $\omega<1/2$). By the definition of $\omega$, the checkpoint count therefore satisfies
\begin{equation*}
    |\mathcal C(\tilde x)|\le\frac{4}{\omega}\cdot\bigl(1+\ceil{\log_2(m\cdot|\mathcal F|)}\bigr)=\frac{|\mathcal F|}{4}.
\end{equation*}
We write $\mathcal F'=\{(i,S):0<x'_i(S)<1\}$ for the fractional support of the basic optimum $x'$ of $\mathrm{LP}(\tilde x)$. 
Once all integral entries of $x'$ are fixed, the coordinates in $\mathcal F'$ still form a basic feasible solution, so $|\mathcal F'|$ is at most the number of tight constraints involving $\mathcal F'$. Since every variable in $\mathcal F'$ has $x'_i(S)>0$ (so $x'_i(S)\ge0$ is never tight) and no constraint $x'_i(S)=\tilde x_i(S)$ involves $\mathcal F'$, only two types of tight constraints remain: checkpoint constraints (at most $|\mathcal C(\tilde x)|\le|\mathcal F|/4$) and tight agent constraints $\sum_{S\in\mathcal B_i}x'_i(S)=1$.
Furthermore, any tight agent constraint containing a variable in $\mathcal F'$ cannot have an entry equal to $1$, so it must contain at least two variables in $\mathcal F'$ to sum to $1$. Since agent constraints are disjoint across agents, at most $|\mathcal F'|/2$ such constraints exist. Therefore, $|\mathcal F'|\le|\mathcal F|/4+|\mathcal F'|/2$, which implies $|\mathcal F'|\le|\mathcal F|/2$.

Since $n$ is sufficiently large, $m\le n^2$ and the initial fractional support satisfies $|\mathcal F|\le n+m\le 2n^2\le 2^{2L+1}$. Because $|\mathcal F|$ halves in every iteration with $\omega<1/2$, the definition of $\omega$ yields $\omega\cdot|\mathcal F|\le 80L$ throughout Step~1, and Step~1 performs at most $2L+1$ such iterations. In each of them, reducing $x$ to $\tilde x$ decreases total weight by at most $\omega\cdot|\mathcal F|\le 80L$, and because $\tilde x$ is feasible for $\mathrm{LP}(\tilde x)$, the optimum $x'$ loses no further weight. If an iteration reaches $\omega\ge1/2$, setting all entries in $\mathcal F$ to $0$ removes $|\mathcal F|\le 2\omega\cdot|\mathcal F|\le 160L$ entries, each of weight below $1$, and terminates Step~1. Upon termination, $x$ is integral with $\sum_{S\in\mathcal B_i}x_i(S)\in\{0,1\}$ for every $i\in N$, so total weight starts at $n$ and ends at $n-|N^-|$, which implies that
\begin{equation*}
    |N^-|\le(2L+1)\cdot 80L+160L<768L^2=R.
\end{equation*}
After fixing the integral coordinates, $\mathrm{LP}(\tilde x)$ has at most $|\mathcal F|\le n+m=O(n)$ free variables and integer constraint coefficients of magnitude at most $m$. The common denominator of the right-hand sides divides the previous common denominator of $x$ times $|\mathcal F|$, since the denominator of $\omega$ divides $|\mathcal F|$. By Cramer's rule and Hadamard's bound, a basis determinant increases the denominator's bit length by at most $O(n\log n)$ per iteration. Since all weights lie in $[0,1]$ and there are $O(L)$ iterations, the bit lengths and the running time of Step~1 are polynomial.
\end{proof}

\begin{lemma}\label{lem:slack-rounding}
If $\Delta\ge2R$, Algorithm~\ref{alg:slack} computes an allocation $\bX$ with $v_i(X_i)\ge\sigma$ for every $i\in N$.
\end{lemma}
\begin{proof}
By Lemmas~\ref{lem:prefix-load} and~\ref{lem:step1-rounding}, Step~1 terminates in polynomial time with an integral weight vector $x$ that preserves every prefix bound $\sum_{i\in N}\sum_{S\in\mathcal B_i}x_i(S)\cdot|S\cap M[j]|\le\lambda\cdot j$ ($j\in[m]$) and leaves $|N^-|\le R$ agents unserved.
In Step~2, the preserved prefix bounds ensure that the $k$-th request in deadline order, say with deadline $j$, satisfies $k\le\lambda\cdot j$ and hence $\ceil{k/\lambda}\le j$. Since $\lambda\le1$, the indices $\ceil{k/\lambda}$ are strictly increasing, so Step~2 assigns distinct items satisfying every request. Every $i\in N\setminus N^-$ therefore receives a bundle $X_i$ of value at least $\sigma$, and each prefix $M[j]$ ($j\in[m]$) has at most $\floor{\lambda\cdot j}$ assigned items, leaving $|U\cap M[j]|\ge(1-\lambda)\cdot j$. The value of $U$ is a weighted sum of these prefix counts. For any nonincreasing sequence $s_1\ge\dots\ge s_\ell\ge s_{\ell+1}=0$ and any subset $S\subseteq[\ell]$, we have
\begin{equation}\label{eq:prefixvalue}
    \sum_{j\in S}s_j=\sum_{j=1}^{\ell}|S\cap[j]|\cdot(s_j-s_{j+1}).
\end{equation}
For each $i\in N^-$, we take $s_j=v_i(e_j)$ for $j\in[m]$ and apply~\eqref{eq:prefixvalue} to the positions of goods in $U$. Since $s_j-s_{j+1}\ge 0$, the prefix bound gives $v_i(U)\ge(1-\lambda)\cdot v_i(M)=\Delta$.

In each iteration of Step~3, the chosen prefix $B'\subseteq U$ has value below $\sigma$ for all waiting agents before its last item is added, and every item has value at most $1$, so removing $B'$ lowers $v_i(U)$ by less than $\sigma+1<2$ for every agent in $N^-$. Since $|N^-|\le R$ initially and $\Delta\ge 2R$, every waiting agent has value at least $\Delta-2(R-1)\ge 2>\sigma$ in $U$ at each step, so Step~3 serves all of $N^-$ with value at least $\sigma$. Finally, the leftover items in $U$ are added to $X_1$ without decreasing $v_1(X_1)$, and Steps~2 and~3 also run in polynomial time.
\end{proof}

\subsection{Small Slack: Dual Characterization}\label{subsec:valuable-prefix}

In the following, we study the small-slack case $\Delta < 2R$. In this case, the LP leaves too little unused capacity for Algorithm~\ref{alg:slack} to guarantee a value of at least $\sigma$ to every agent.
Intuitively, if $n$ agents cannot even save $O(\log^2 n)$ bundles' worth of items by trading across prefixes, then almost all agents must depend on the same set of large items, while the remaining smaller items at the back hold very little value. We make this intuition precise by looking at the dual of the LP. The dual gives us item prices that split the agents into \emph{ordinary} and \emph{reserved} agents (Definition~\ref{def:ordinary-reserved}) and identify a single prefix $M_1$ of valuable items that holds almost all of the value for every ordinary agent (Lemma~\ref{lem:valuable-prefix}).

In the dual LP, we introduce one variable $\hat{\zeta}_i\in\mathbb R$ for each agent constraint $\sum_{S\in\mathcal B_i} x_i(S)=1$, and one nonnegative variable $\hat{\alpha}_j\ge 0$ for each prefix constraint $j\in[m]$. To make the numbers easier to interpret, we multiply every dual variable by $n$, setting $\zeta_i = n\cdot \hat{\zeta}_i$ and $\alpha_j = n\cdot \hat{\alpha}_j \ge 0$.
For each item $e_k\in M$, we define its \emph{price} as $p(e_k) = \sum_{j=k}^{m} \alpha_j$. 
Notice that the prices are nonincreasing and the sum of all item prices equals $n$:
\begin{equation*}
    p(M) = \sum_{k=1}^m p(e_k) = \sum_{k=1}^m \sum_{j=k}^m \alpha_j = \sum_{j=1}^m \sum_{k=1}^j \alpha_j = \sum_{j=1}^m j\cdot \alpha_j = n,
\end{equation*}
where the last equality holds since in the unscaled dual, the constraint for the primal variable $\lambda$ is $\sum_{j=1}^m j\cdot \hat{\alpha}_j = 1$.
Furthermore, for any bundle $S\subseteq M$, writing $p(S) = \sum_{e\in S} p(e)$, we have
\begin{equation*}
    \sum_{j=1}^m |S\cap M[j]|\cdot \alpha_j
    = \sum_{e_k\in S} \sum_{j=k}^m \alpha_j
    = \sum_{e_k\in S} p(e_k) = p(S).
\end{equation*}
Therefore, the scaled dual constraint for the primal variable $x_i(S)$ is simply $\zeta_i \le p(S)$.

Putting these three facts together, the scaled dual LP is
\begin{align*}
    \text{maximize}\quad & \textstyle \sum_{i\in N}\zeta_i \\
    \text{subject to}\quad
    & p(e_1)\ge\dots\ge p(e_m)\ge 0, \\
    & \textstyle  \sum_{e\in M}p(e)=n, \\
    & \zeta_i\le p(S) && \forall i\in N,\ S\in\mathcal B_i.
\end{align*}

At an optimal dual solution, we have $\zeta_i = \min_{S\in\mathcal B_i}\{p(S)\} \geq 0$ for every $i\in N$. On the other hand, the family $\mathcal B_i$ contains all $n$ disjoint bundles of agent $i$'s MMS partition $\bP^i$. Since these $n$ bundles partition $M$, their prices sum to $p(M) = n$, so their average price is $1$. Thus, we have $\zeta_i \le 1$.
By strong duality, we have $\sum_{i\in N} \zeta_i = n\cdot \lambda$, which implies
\begin{equation*}
    \sum_{i\in N}\left(1-\zeta_i\right) = n\cdot\left(1-\lambda\right) = \Delta.
\end{equation*}
Note that every term $1-\zeta_i$ is nonnegative (since $\zeta_i \le 1$), and their sum is $\Delta < 2R = O(L^2)$. Therefore, almost every agent must have $1 - \zeta_i$ very close to $0$ (that is, $\zeta_i$ very close to $1$).

\begin{definition}\label{def:ordinary-reserved}
We partition $N$ into the \emph{ordinary agents} $N_o$ and the \emph{reserved agents} $N_r$ by
\begin{equation*}
    N_o = \left\{i\in N : n\cdot\left(1-\zeta_i\right) \le L^4\right\}
    \qquad\text{and}\qquad
    N_r = N \setminus N_o.
\end{equation*}
\end{definition}

Since every agent $i\in N_r$ satisfies $n\cdot(1-\zeta_i) > L^4$, we have
\begin{equation*}
    |N_r|\cdot L^4 \leq \sum_{i\in N_r} n\cdot\left(1-\zeta_i\right) \le \sum_{i\in N} n\cdot\left(1-\zeta_i\right) = n\cdot\Delta < 2n\cdot R = 1536 n\cdot L^2.
\end{equation*}
Dividing both sides by $L^4$ yields $|N_r| \le 1536 n / L^2 = O(n/L^2) = o(n)$.

\begin{lemma}\label{lem:valuable-prefix}
If $\Delta < 2R$, then we can compute in polynomial time a nonempty prefix $M_1\subseteq M$ such that every $i\in N_o$ satisfies $v_i(e) > 1-\sigma$ for all $e\in M_1$ and $v_i(M\setminus M_1) = O(L^4)$.
\end{lemma}
\begin{proof}
Fix any agent $i\in N$ and any bundle $B\in\bP^i$. Recall how we formed the reduced MMS bundles from $B$ in Subsection~\ref{subsec:lp-framework}: we take all items $e\in B$ with $v_i(e)\le 1-\sigma$ in the common order and greedily partition them from left to right into consecutive segments $B'_1, B'_2, \dots, B'_t$, where each segment has value at most $1-\sigma$ and is maximal.
The total number of segments is at most $1 + \ceil{2/(1-\sigma)}$.
Now look at any single $B'\subseteq B$, where $B\setminus B'$ belongs to $\mathcal B_i$. By the dual constraint for $B\setminus B'$, we have $p(B\setminus B') \ge \zeta_i$, which gives $p(B') \le p(B) - \zeta_i$.
Note that $p(B) - \zeta_i \ge 0$ because the whole bundle $B$ also belongs to $\mathcal B_i$. Summing $p(B') \le p(B) - \zeta_i$ over all segments inside $B$ gives
\begin{equation*}
    \sum_{e\in B:\, v_i(e)\le 1-\sigma} p(e)
    \le \left(1 + \ceil{2/(1-\sigma)}\right)\cdot\left(p(B) - \zeta_i\right).
\end{equation*}
Finally, summing this inequality over all $n$ bundles $B\in\bP^i$ yields
\begin{equation}\label{eq:weakprices}
    \sum_{e\in M:\, v_i(e)\le 1-\sigma} p(e)
    \le \left(1 + \ceil{2/(1-\sigma)}\right)\cdot n\cdot\left(1-\zeta_i\right)
    \qquad \forall i\in N.
\end{equation}

Let $M_2$ be the set of items $e$ for which there exists an ordinary agent $i\in N_o$ such that $v_i(e) \leq 1-\sigma$.
Set $M_1 = M\setminus M_2$.
From this partition, two properties follow immediately:
\begin{itemize}
    \item Every item in $M_1$ has value strictly greater than $1-\sigma$ for every ordinary agent $i\in N_o$.
    \item There exists an ordinary agent $i^* \in N_o$ that values every item in $M_2$ at most $1-\sigma$. Applying~\eqref{eq:weakprices} to $i^*\in N_o$ and using Definition~\ref{def:ordinary-reserved} ($n\cdot(1-\zeta_{i^*}) \le L^4$), we obtain
    \begin{equation*}
        p(M_2) \le \left(1 + \ceil{2/(1-\sigma)}\right)\cdot L^4.
    \end{equation*}
\end{itemize}

Finally, we convert the price bound on $M_2$ into a value bound.
Fix any ordinary agent $i\in N_o$. We scan $M_2$ from left to right and repeatedly cut off the shortest remaining prefix of value at least $\sigma$ to agent $i$, stopping when the remaining items of $M_2$ have total value less than $\sigma$. 
Suppose this produces $s+1$ blocks.
Each of the first $s$ blocks belongs to the candidate family $\mathcal B_i$, so its dual price is at least $\zeta_i \ge 1 - L^4/n \ge 1/2$. Because the blocks are disjoint subsets of $M_2$, summing their prices gives $p(M_2) \geq s/2$.
By our construction, every block has value at most $1+\sigma\leq 2$.
Summing over all $s+1 \le 2p(M_2)+1$ blocks gives $v_i(M_2) \le (2p(M_2)+1)\cdot 2 = O(L^4)$.
In particular, $v_i(M_2) = o(n)$, so $v_i(M_1) = n - v_i(M_2) > 0$, which confirms that $M_1$ is nonempty. Finally, we can find $i^*$ and $M_1$ in polynomial time simply by scanning the item values of all agents in $N_o$.
\end{proof}

We set the constant $\kappa = \ceil{1/(1-\sigma)}$. In any MMS bundle $P_j^i$ of an ordinary agent $i\in N_o$, every item from $M_1$ has value strictly greater than $1-\sigma$. Since the total value of $P_j^i$ is $1$, we have $|P_j^i\cap M_1| < \kappa$ for every $i\in N_o$ and $j\in[n]$.

\paragraph{Summary.}
We summarize the properties we have established when $\Delta < 2R$. We work with the normalized IDO instance on $M$ with $m=O(n)$ goods and MMS partitions $\{\bP^i\}_{i\in N}$ from Section~\ref{subsec:lp-framework}, together with the agent partition and prefix $M_1$ computed in polynomial time above:
\begin{enumerate}
    \renewcommand{\labelenumi}{(\arabic{enumi})}
    \item\label{prop:reserved-small}
    The agents $N = N_o\cup N_r$ are partitioned into ordinary agents $N_o$ and reserved agents $N_r$, where we have $|N_r| = O(n/L^2) = o(n)$.
    \item\label{prop:common-prefix}
    The prefix $M_1\subseteq M$ is nonempty, and $v_i(M_2) = O(L^4)$ for every $i\in N_o$, where $M_2 = M\setminus M_1$.
    \item\label{prop:prefix-count}
    For the constant $\kappa = \ceil{1/(1-\sigma)}$, we have $|P_j^i\cap M_1| < \kappa$ for every $i\in N_o$ and $j\in[n]$.
\end{enumerate}

\subsection{Small Slack: Selecting Bundles}\label{sec:prefix-allocation}

In this subsection and the next, we use Properties~(\ref{prop:reserved-small})--(\ref{prop:prefix-count}) to compute a $\beta$-MMS allocation in polynomial time (Theorem~\ref{thm:common-prefix-allocation}). Notice first that summing Property~(\ref{prop:prefix-count}) over all $n$ bundles of any single ordinary agent's MMS partition gives $|M_1| \le \kappa\cdot n$. Throughout both subsections, $\beta\in(0,1)$ and $\kappa$ are fixed constants, and $n$ is sufficiently large. 
By Property~(\ref{prop:common-prefix}), we also have
\begin{equation*}
    T = \max\left\{1,\left\lceil\max_{i\in N_o}\{v_i(M_2)\}\right\rceil\right\} = O(L^4) = n^{o(1)}.
\end{equation*}

In the following, we divide $M_1$ into a constant number of intervals, group almost all ordinary agents, and select $n$ integral bundles. We then assign actual items in Section~\ref{subsec:completing-allocation}.

\subsubsection{Initial Intervals}\label{subsubsec:initial-intervals}

We first fix a rational rounding step $d_1 > 0$ on $M_1$ small enough that $(\kappa + 1)\cdot d_1 < 1 - \beta$.
For every ordinary agent $i\in N_o$ and every item $e\in M_1$, we round $v_i(e)$ down to the nearest multiple of $d_1$. Since an ordinary agent's MMS bundle has fewer than $\kappa$ items in $M_1$, rounding down loses strictly less than $\kappa\cdot d_1$ across those items.
Now look at one ordinary agent $i$'s $d_1$-rounded values as we scan $M_1$ from left to right. Because the instance is IDO, her rounded values are nonincreasing and lie in $[0,1]$. Each time her rounded value drops between two neighboring positions in $M_1$, it drops by at least $d_1$. 
Thus, she has at most $\floor{1/d_1}$ drop positions across all of $M_1$. We place a \emph{mark} at every drop position of every ordinary agent $i\in N_o$. Across all $|N_o|\le n$ ordinary agents combined, there are at most $n\cdot\floor{1/d_1}$ marks in total. We order all marks by position along $M_1$ and place a cut at every $\ceil{n/10}$-th mark. This places at most $\floor{10/d_1}$ cuts.

The marks can leave a long stretch of items between two cuts (which leads to some difficulties in our later analysis).
We therefore add cuts at regular distances.
Specifically, we divide $M_1$ into blocks of $\ceil{d_1\cdot |M_1|}$ consecutive items, allowing the last block to be shorter. We place a cut between each pair of consecutive blocks. Since $|M_1|/\ceil{d_1\cdot |M_1|}\le 1/d_1$, we add at most $\floor{1/d_1}$ cuts. Together the cuts give at most $\floor{11/d_1}+1$ nonempty \emph{initial intervals}.

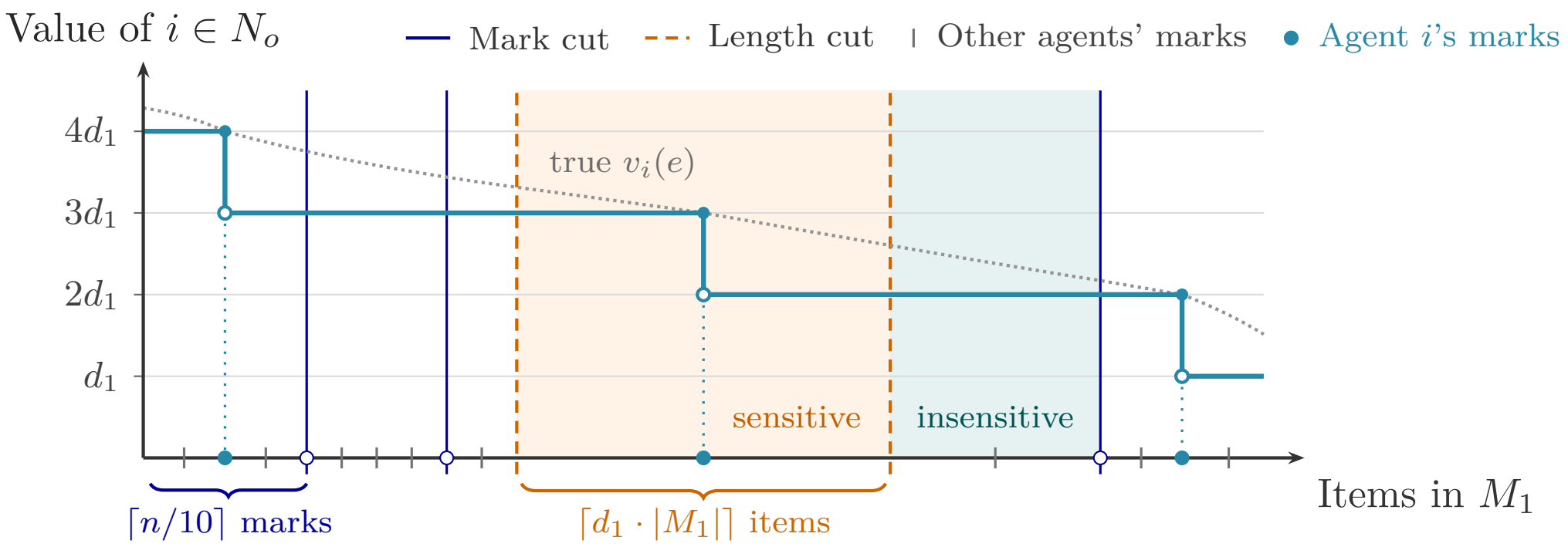
\begin{figure}[htbp]
  \centering
  \begin{tikzpicture}[
    x=1cm, y=1cm, font=\footnotesize,
    every node/.style={inner sep=1.5pt}
  ]
    \fill[orange!10] (3.20, 0) rectangle (6.40, 3.15);
    \fill[teal!10]   (6.40, 0) rectangle (8.20, 3.15);

    \foreach \y/\lab in {0.70/$d_1$, 1.40/$2d_1$, 2.10/$3d_1$, 2.80/$4d_1$}{
      \draw[black!14, thin] (0, \y) -- (9.60, \y);
      \draw[black!60] (-0.08, \y) -- (0, \y);
      \node[anchor=east, text=black!75] at (-0.14, \y) {\lab};
    }

    \foreach \xc in {1.40, 2.60, 8.20}{
      \draw[blue!60!black, semithick] (\xc, -0.14) -- (\xc, 3.15);
    }
    \foreach \xc in {3.20, 6.40}{
      \draw[orange!80!black, densely dashed, thick] (\xc, -0.14) -- (\xc, 3.15);
    }

    \draw[black!42, thick, densely dotted]
      (0.00, 3.00)
      .. controls (0.32, 2.95) and (0.55, 2.87) .. (0.70, 2.80)
      .. controls (1.55, 2.56) and (2.60, 2.40) .. (3.20, 2.32)
      .. controls (3.95, 2.22) and (4.45, 2.15) .. (4.80, 2.10)
      .. controls (5.95, 1.90) and (7.30, 1.66) .. (8.20, 1.52)
      .. controls (8.50, 1.47) and (8.72, 1.43) .. (8.90, 1.40)
      .. controls (9.22, 1.28) and (9.45, 1.15) .. (9.60, 1.06);

    \draw[cyan!60!black, very thick]
      (0.00, 2.80) -- (0.70, 2.80)
      -- (0.70, 2.10) -- (4.80, 2.10)
      -- (4.80, 1.40) -- (8.90, 1.40)
      -- (8.90, 0.70) -- (9.60, 0.70);

    \foreach \xd/\ytop/\ybot in {0.70/2.80/2.10, 4.80/2.10/1.40, 8.90/1.40/0.70}{
      \draw[cyan!60!black, dotted, semithick] (\xd, \ybot) -- (\xd, 0);
      \fill[cyan!60!black] (\xd, \ytop) circle (1.5pt);
      \fill[white] (\xd, \ybot) circle (1.5pt);
      \draw[cyan!60!black, thick] (\xd, \ybot) circle (1.5pt);
    }

    \draw[-{Stealth[length=4pt]}, thick, black!80] (0, 0) -- (9.95, 0);
    \draw[-{Stealth[length=4pt]}, thick, black!80] (0, 0) -- (0, 3.40);
    \node[anchor=south, text=black!85] at (0, 3.46) {\small Value of $i\in N_o$};

    \foreach \xm in {0.35, 1.05, 1.70, 2.00, 2.30, 2.90, 7.30, 8.55, 9.30}{
      \draw[black!55, semithick] (\xm, -0.09) -- (\xm, 0.09);
    }
    \foreach \xm in {0.70, 4.80, 8.90}{
      \fill[cyan!60!black] (\xm, 0) circle (1.8pt);
    }
    \foreach \xc in {1.40, 2.60, 8.20}{
      \filldraw[fill=white, draw=blue!60!black, thin] (\xc, 0) circle (1.6pt);
    }

    \node[anchor=west, text=black!58] at (3.42, 2.52) {\scriptsize true $v_i(e)$};

    \node[text=orange!75!black] at (5.60, 0.35) {\scriptsize sensitive};
    \node[text=teal!65!black]   at (7.30, 0.35) {\scriptsize insensitive};

    \draw[decorate, decoration={brace, amplitude=3.2pt, mirror}, blue!60!black, thick]
      (0.06, -0.22) -- (1.40, -0.22)
      node[midway, below=4pt, text=blue!60!black] {\scriptsize $\ceil{n/10}$ marks};

    \draw[decorate, decoration={brace, amplitude=3.5pt, mirror}, orange!80!black, thick]
      (3.24, -0.22) -- (6.36, -0.22)
      node[midway, below=4pt, text=orange!80!black] {\scriptsize $\ceil{d_1\cdot|M_1|}$ items};

    \node[anchor=north west, text=black!80] at (10.00, -0.12)
      {\small Items in $M_1$ };

    \draw[blue!60!black, thick] (2.25, 3.60) -- (2.63, 3.60);
    \node[anchor=west, text=black!80] at (2.73, 3.60) {\scriptsize Mark cut};

    \draw[orange!80!black, densely dashed, thick] (4.30, 3.60) -- (4.68, 3.60);
    \node[anchor=west, text=black!80] at (4.78, 3.60) {\scriptsize Length cut};

    \draw[black!55, semithick] (6.60, 3.52) -- (6.60, 3.68);
    \node[anchor=west, text=black!75] (other-marks) at (6.74, 3.60) {\scriptsize Other agents' marks};

    \fill[cyan!60!black] ([xshift=9pt]other-marks.east) circle (1.8pt);
    \node[anchor=west, text=cyan!60!black] at ([xshift=14pt]other-marks.east) {\scriptsize Agent $i$'s marks};
  \end{tikzpicture}
  \caption{Initial intervals on $M_1$ formed by mark cuts and length cuts.}
  \label{fig:initial-intervals}
\end{figure}

We say that an ordinary agent $i\in N_o$ is \emph{sensitive} to an interval if her $d_1$-rounded value drops strictly inside that interval, and \emph{insensitive} to it otherwise. 
By the cuts at the marked positions, strictly fewer than $\ceil{n/10}$ marks lie inside any initial interval, so at most $n/10 + 1$ ordinary agents are sensitive to it. 
Intuitively, on any interval to which agent $i$ is insensitive, her $d_1$-rounded value is completely flat (constant) across the whole interval, so swapping any item in that interval for another changes her true value by strictly less than $d_1$.
See Figure~\ref{fig:initial-intervals} for an illustration.

\subsubsection{Refined Intervals and Aligned Groups}
\label{subsubsec:local-records}

We next fix two small positive rationals for $M_2$: a value threshold $\tau > 0$ and a rounding step $d_2 > 0$, chosen small enough that $\kappa\cdot d_1 + \tau + d_2/\tau < 1 - \beta$.
This is possible because $(\kappa+1)\cdot d_1 < 1 - \beta$.
We will choose an integer $\gamma=n^{o(1)}$ with $\sqrt \gamma\ge 2L$ before refining the initial intervals. We give its formula in Appendix~\ref{app:bundle-selection}. For each ordinary agent $i\in N_o$, we define her \emph{local record} to be the following collection of rounded value lists:
\begin{enumerate}
    \item For each initial interval, her $d_1$-rounded values on its first $2\gamma\cdot(1+\floor{1/d_1})$ local positions (or on the whole interval if it has fewer than $2\gamma\cdot(1+\floor{1/d_1})$ items).
    \item On $M_2$, the list of all items that agent $i$ values at least $\tau$, together with their values rounded down to integer multiples of $d_2$. Because the instance is IDO, these items form a prefix of $M_2$.
\end{enumerate}
Let us count how many different local records can exist. Look at any single list inside a local record:
\begin{itemize}
    \item On an initial interval, the list is nonincreasing, has length at most $2\gamma\cdot(1+\floor{1/d_1}) = O(\gamma)$, and takes values from the $1+\floor{1/d_1}$ multiples of $d_1$ in $[0,1]$. Because the list is sorted from largest to smallest, we can reconstruct the entire list (and its length) simply by stating how many items sit at each of the $1+\floor{1/d_1}$ levels. Each count is an integer between $0$ and $O(\gamma)$, so there are at most $(O(\gamma)+1)^{1+\floor{1/d_1}}$ possible lists for one initial interval.
    \item On $M_2$, every ordinary agent has $v_i(M_2) \le T$, so she has at most $T/\tau$ items worth at least $\tau$. Thus, her $M_2$ list has length at most $T/\tau = O(T)$, and takes values from at most $1+\floor{1/d_2}$ multiples of $d_2$. Counting how many items sit at each level shows there are at most $(O(T)+1)^{1+\floor{1/d_2}}$ possible lists for $M_2$.
\end{itemize}
Because there are at most $\floor{11/d_1}+1 = O(1)$ initial intervals, multiplying these counts across all lists shows that the total number of distinct local records is at most $(\gamma + T + 2)^{O(1)} = n^{o(1)}$. 
We partition the ordinary agents $N_o$ into classes of agents who have the same local record, and we pick one \emph{largest} class $N_a \subseteq N_o$.
By Property~(\ref{prop:reserved-small}), there are $|N_o| = n - o(n)$ ordinary agents. By the pigeonhole principle, the largest class $N_a$ has size at least $|N_o|/n^{o(1)} = n^{1-o(1)}$. This property will be used in the analysis of Section~\ref{subsubsec:suffix-sharing}.

Because every agent in $N_a$ has the same list for $M_2$, all agents in $N_a$ agree on the same prefix $M'_2 \subseteq M_2$ of items worth at least $\tau$, and on the same remaining items $M''_2 = M_2\setminus M'_2$ of items worth strictly less than $\tau$. Inside $N_a$, we pick one agent $a\in N_a$ who has the smallest total value on $M''_2$, and we call $a$ the \emph{reference agent}. Intuitively, every agent in $N_a$ agrees closely with $a$ on the front of every initial interval and on $M'_2$, while valuing $M''_2$ at least as much as $a$ does, so any bundle from $\bP^a$ has nearly the same value to every agent in $N_a$ as it does to $a$.

Given any refinement $I_1,\ldots,I_q$ of the initial intervals and any $\ell\in[q]$, we call the local positions $1,2,4,\ldots$ up to $|I_\ell|$, together with the last position $|I_\ell|$, the \emph{checkpoints} of $I_\ell$. We call a set of ordinary agents \emph{aligned} if their $d_1$-rounded values agree at every checkpoint of every refined interval.

\begin{lemma}\label{lem:representative}
For any positive integer $\gamma = n^{o(1)}$, in polynomial time, we can refine the initial intervals into $I_1,\dots,I_q$, set aside a subset $N_s\subseteq N$, and partition $N\setminus N_s$ into aligned groups $\{N_g\}$\footnote{Note that each $N_g$ is the set of agents in group $g$, and $\{N_g\}$ denotes the entire partition of $N\setminus N_s$.} with the following properties:
\begin{enumerate}
    \item We have $q \le \floor{12/d_1} + 1$. Each $I_\ell$ contains at most $\ceil{d_1\cdot |M_1|}$ items, and at most $n/10 + 1$ ordinary agents are sensitive to it.
    \item On every interval $I_\ell$, either all agents in $N_a$ share one constant $d_1$-rounded value across all of $I_\ell$, or they agree with $a$'s highest $d_1$-rounded value in $I_\ell$ in the first $2\gamma$ local positions of $I_\ell$.
    \item $N_s$ contains $\gamma$ members of $N_a$ and all of $N_r$, and satisfies $n/L \le |N_s| = o(n)$.
    \item There are at most $G = \bigl(2 + \ceil{\log_2(\kappa\cdot n)}\bigr)^{q\cdot(1+\floor{1/d_1})} = n^{o(1)}$ aligned groups $\{N_g\}$, and every group satisfies $|N_g| \ge \sqrt n$.
\end{enumerate}
\end{lemma}
\begin{proof}
We first refine the initial intervals around the reference agent $a$.
Take each initial interval one at a time and look at the $d_1$-rounded values of agent $a$ on it:
\begin{itemize}
    \item \emph{Case A: The whole initial interval fits inside $a$'s local record.} Then every agent in $N_a$ has the exact same $d_1$-rounded values as $a$ on the entire interval. We cut this initial interval at every position where $a$'s $d_1$-rounded value drops. Every resulting sub-interval now has one constant $d_1$-rounded value shared by all of $N_a$.
    \item \emph{Case B: The initial interval is longer than $2\gamma\cdot(1+\floor{1/d_1})$.} Look at its recorded prefix of length $2\gamma\cdot(1+\floor{1/d_1})$. Because $a$'s $d_1$-rounded values are nonincreasing and take at most $1+\floor{1/d_1}$ distinct values, this prefix splits into at most $1+\floor{1/d_1}$ flat runs of equal rounded value. By the pigeonhole principle, at least one flat run has length at least $2\gamma$. We cut at every value drop of $a$ up to the start of this long run, and make no cuts after it. Every sub-interval created before the long run is completely flat for all of $N_a$; and the final sub-interval starts with this flat run of length at least $2\gamma$, so every agent in $N_a$ agrees with $a$'s highest $d_1$-rounded value on this sub-interval across its entire first $2\gamma$ positions.
\end{itemize}

Every cut added in this refinement occurs at a position where $a$'s $d_1$-rounded value drops. Because $a$'s $d_1$-rounded values only decrease across the whole prefix $M_1$ (from at most $1$ to at least $0$), agent $a$ has at most $\floor{1/d_1}$ drops across all of $M_1$ combined. Thus, we add at most $\floor{1/d_1}$ cuts in total across all initial intervals, giving $q \le \floor{11/d_1} + 1 + \floor{1/d_1}\leq \floor{12/d_1} + 1$. Each refined interval lies in an initial interval and hence contains at most $\ceil{d_1\cdot |M_1|}$ items. Since at most $n/10 + 1$ ordinary agents are sensitive to each initial interval, the same applies to its refinements. This establishes properties~(1) and~(2).

\smallskip

Next, we introduce the partition of agents into groups.
Start forming $N_s$ by putting in $\gamma$ arbitrary agents from $N_a$ together with all reserved agents $N_r$. If the pool has fewer than $\ceil{n/L}$ agents so far, add arbitrary ordinary agents until $|N_s| = \ceil{n/L}$.

For every remaining agent $i\in N\setminus N_s$, we create an \emph{alignment label} by recording her $d_1$-rounded values at the checkpoints of each refined interval $I_\ell$. The list of values is nonincreasing, so it is determined by the number of checkpoints at each of the $1+\floor{1/d_1}$ possible rounded value levels. Since $|I_\ell| \le |M_1| \le \kappa\cdot n$, there are at most $1 + \ceil{\log_2(\kappa\cdot n)}$ checkpoints per interval. Each count ranges from zero to the number of checkpoints, giving at most $2 + \ceil{\log_2(\kappa\cdot n)}$ choices per interval and value level. Therefore, there are at most $G$ distinct alignment labels. We group the agents in $N\setminus N_s$ by identical alignment labels; by definition, every resulting group is aligned.

Finally, move every aligned group of size strictly smaller than $\sqrt n$ into $N_s$. This adds at most $G\cdot\sqrt n = o(n)$ agents to $N_s$. Combining this with $\gamma = n^{o(1)}$, $|N_r| = O(n/L^2) = o(n)$, and the initial padding to $\ceil{n/L}$, the final set-aside pool satisfies $n/L \le |N_s| = o(n)$, and every remaining aligned group $N_g$ has size $|N_g| \ge \sqrt n$. Every step runs in polynomial time.
\end{proof}

\subsubsection{Selection of Bundles}\label{subsubsec:quotas}

We call an MMS bundle $P_j^i$ \emph{eligible} for an ordinary agent $i\in N_o$ if $v_i(P_j^i\cap M_2) \le d_1$. Note that every eligible bundle satisfies $v_i(P_j^i\cap M_1) \ge 1 - d_1$, which still leaves value strictly above $1 - (\kappa+1)\cdot d_1 > \beta$ after the $d_1$-rounding loss of less than $\kappa\cdot d_1$ on $M_1$, so its recipient needs no items from $M_2$.
For each aligned group $N_g$ from Lemma~\ref{lem:representative}, pick one arbitrary member $a_g\in N_g$ as the group's \emph{representative}, and give every bundle in $a_g$'s MMS partition $\bP^{a_g} = (P_1^{a_g},\dots,P_n^{a_g})$ the fractional weight
\begin{equation*}
    \phi_g = \frac{|N_g|}{|N\setminus N_s|} = \frac{|N_g|}{n - |N_s|}.
\end{equation*}
Notice two immediate consequences of this weighting:
\begin{enumerate}
    \item Because the groups $N_g$ partition $N\setminus N_s$, we have $\sum_g \phi_g = 1$. Since each partition $\bP^{a_g}$ uses every item of every interval $I_\ell$ exactly once, the total fractional number of items used in $I_\ell$ across all groups is $\sum_g \phi_g\cdot |I_\ell| = |I_\ell|$.
    \item Because $\bP^{a_g}$ has $n$ bundles, the total fractional number of bundles assigned to group $N_g$ is $\phi_g\cdot n = |N_g|\cdot\frac{n}{n-|N_s|} = |N_g| + \frac{|N_s|\cdot |N_g|}{n-|N_s|}$. Since $|N_s|\ge n/L$ and $|N_g|\ge\sqrt n$, every group receives at least $\sqrt n/L$ extra fractional bundles beyond the $|N_g|$ bundles its members need.
\end{enumerate}

Next, we define how a bundle of representative $a_g$ specifies item requests for any member $i\in N_g$:
\begin{itemize}
    \item On an interval $I_\ell$ to which $a_g$ is \emph{insensitive}, every member $i\in N_g$ shares the same constant $d_1$-rounded value across $I_\ell$ (because their rounded values agree at the first and last checkpoints). 
    For every item in the representative's bundle that belongs to $I_\ell$, we create one request for an item from $I_\ell$. Any available item in $I_\ell$ satisfies the request, so it has \emph{no deadline}.
    \item On an interval $I_\ell$ to which $a_g$ is \emph{sensitive}, an item at local position $r$ in a bundle of $\bP^{a_g}$ generates a \emph{sensitive request} whose deadline is the last checkpoint at or before $r$. Every member of $N_g$ agrees with $a_g$ on the rounded value at that checkpoint. Since rounded values are nonincreasing, any item at or before the deadline gives a group member at least the rounded value that $a_g$ assigns to position $r$.
\end{itemize}

For each aligned group $N_g$, we call the $n$ bundles in its representative $a_g$'s MMS partition $\bP^{a_g}$ the \emph{representative bundles} of $N_g$.
Since there are at most $G$ groups, there are at most $n\cdot G$
representative bundles in total.
Recall that $a$ is the reference agent chosen from $N_a$.

\begin{lemma}\label{lem:quotas}
We can in polynomial time select $n-k$ eligible representative bundles and $k\le \sqrt \gamma$ distinct bundles of $\bP^a$ (called \emph{backup bundles}) satisfying:
\begin{enumerate}
    \item The $n$ selected bundles together request exactly $|I_\ell|$ items from every interval $I_\ell$.
    
    \item The selected representative bundles generate strictly fewer than $b/4$ sensitive requests with deadline at most $b$, in every interval $I_\ell$ and every prefix length $b\in[|I_\ell|]$.
    
    \item Each group $N_g$ retains between $\phi_g\cdot n - \sqrt \gamma$ and $\phi_g\cdot n$ representative bundles. Moreover, in every interval $I_\ell$, group $N_g$ retains at most $\phi_g\cdot |I_\ell|$ requests from $I_\ell$, and the sum of its request shortfalls below $\phi_g\cdot |I_\ell|$ across all $q$ intervals is at most $\kappa\cdot\sqrt \gamma$.
\end{enumerate}
\end{lemma}

We select the bundles in two stages, starting from the initial fractional weighting $\phi_g$ on every representative bundle. In the first stage, we discard all ineligible bundles and round the remaining eligible bundles to integral choices using an LP that preserves each group's bundle and interval totals as well as the sensitive deadline loads at all checkpoints. Taking a vertex solution leaves only a small number of fractional bundles, which we also discard. The surviving bundles of weight $1$ form an integral collection of eligible representative bundles that already satisfies the sensitive deadline bound, leaving only a small shortage in the total number of bundles and in the item counts of the intervals.
In the second stage, we repair these shortages so that the final selection has exactly $n$ bundles requesting exactly $\vert{}I_\ell\vert{}$ items from each interval $I_\ell$. Because a bundle from the reference partition $\bP^a$ can request items from several intervals at once, simply adding bundles from $\bP^a$ to fill a shortage in one interval may create too many requests in another. We therefore allow both adding bundles from $\bP^a$ (the backup bundles) and discarding a few more surviving representative bundles, choosing a combination that achieves the exact target counts with the fewest total additions and deletions. Such a repair always exists because $\bP^a$ itself is a complete partition of $M_1$ into $n$ bundles, and it can be found in polynomial time because the number of intervals $q$ and the maximum number of items per bundle $\kappa$ are both constants.
In Appendix~\ref{app:bundle-selection}, we prove that the repair requires few changes and that the selected bundles satisfy all three conclusions of Lemma~\ref{lem:quotas}.

\subsection{Small Slack: Completing the Allocation}\label{subsec:completing-allocation}

We now turn the $n$ abstract bundles from Lemma~\ref{lem:quotas} into a real allocation.

\subsubsection{Backup Agents and Allocation of \texorpdfstring{$M_2$}{}}\label{subsubsec:suffix-sharing}

\begin{figure}[!htbp]
\centering
\begin{tikzpicture}[
    x=1cm, y=1cm, font=\small,
    every node/.style={font=\small},
    group border/.style={draw=blue!45!black, line width=.46pt},
    brace label/.style={font=\small, align=center, text=black!85},
    set brace/.style={decorate, decoration={brace, amplitude=4pt},
        draw=black!75, line width=.45pt}
]
  \fill[blue!4] (0,0) rectangle (8.15,4.00);
  \fill[blue!9] (0,0) rectangle (8.15,.76);
  \fill[teal!9] (8.15,0) rectangle (11.35,4.00);
  \fill[black!10] (11.35,0) rectangle (14.05,4.00);

  \draw[group border] (0,.76) -- (8.15,.76);
  \draw[group border] (0,1.95) -- (8.15,1.95);
  \draw[black!78,line width=.8pt] (8.15,0) -- (8.15,4.00);
  \draw[group border] (11.35,0) -- (11.35,4.00);
  \draw[black!78,line width=.65pt] (0,0) rectangle (14.05,4.00);

  \node[anchor=west,align=left] at (.33,3.05)
    {\textbf{Aligned group} $N_g$\\[-1pt]
     \footnotesize Representative $a_g\in N_g$};
  \node[anchor=west] at (.33,1.33) {\textbf{Aligned group} $N_{g'}$};
  \node[anchor=west,text=black!78,font=\footnotesize] at (.33,.38)
    {Other aligned groups};
  \node[align=center] at (12.70,2.05) {$N_r$\\[-1pt]\footnotesize Reserved agents};

  \draw[violet!70!black,line width=1pt,dash pattern=on 3.6pt off 2.4pt]
    (7.31,2.02) ellipse [x radius=3.72, y radius=1.52];
  \node[text=violet!75!black,align=center] at (6.36,2.77)
    {\textbf{Record class} $N_a$\\[-1pt]
     \footnotesize Reference agent $a\in N_a$};

  \node[draw=orange!68!black,fill=orange!17,text=black!85,
    rounded corners=3pt,line width=.65pt,align=center,
    inner xsep=3pt,minimum width=2.05cm,minimum height=.73cm] at (9.48,2.22)
    {$\boldsymbol{N_b}$\\[-1pt]\footnotesize Backup agents};
  \node[text=violet!75!black,font=\footnotesize] at (9.35,1.45)
    {$|N_a\cap N_s|\ge \gamma$};

  \draw[set brace] (0,4.10) -- (8.15,4.10)
    node[brace label,midway,above=5pt]
      {Agents outside the set-aside pool, $N\setminus N_s$};
  \draw[set brace] (8.15,4.10) -- (14.05,4.10)
    node[brace label,midway,above=5pt] {Set-aside agents, $N_s$};
  \node[font=\small\bfseries] at (7.025,5.05) {All agents, $N$};
  \draw[set brace] (11.35,-.08) -- (0,-.08)
    node[brace label,midway,below=5pt] {Ordinary agents, $N_o$};
\end{tikzpicture}
\caption{Illustration of the agent sets in the small-slack case.}
\label{fig:mms-agent-sets}
\end{figure}
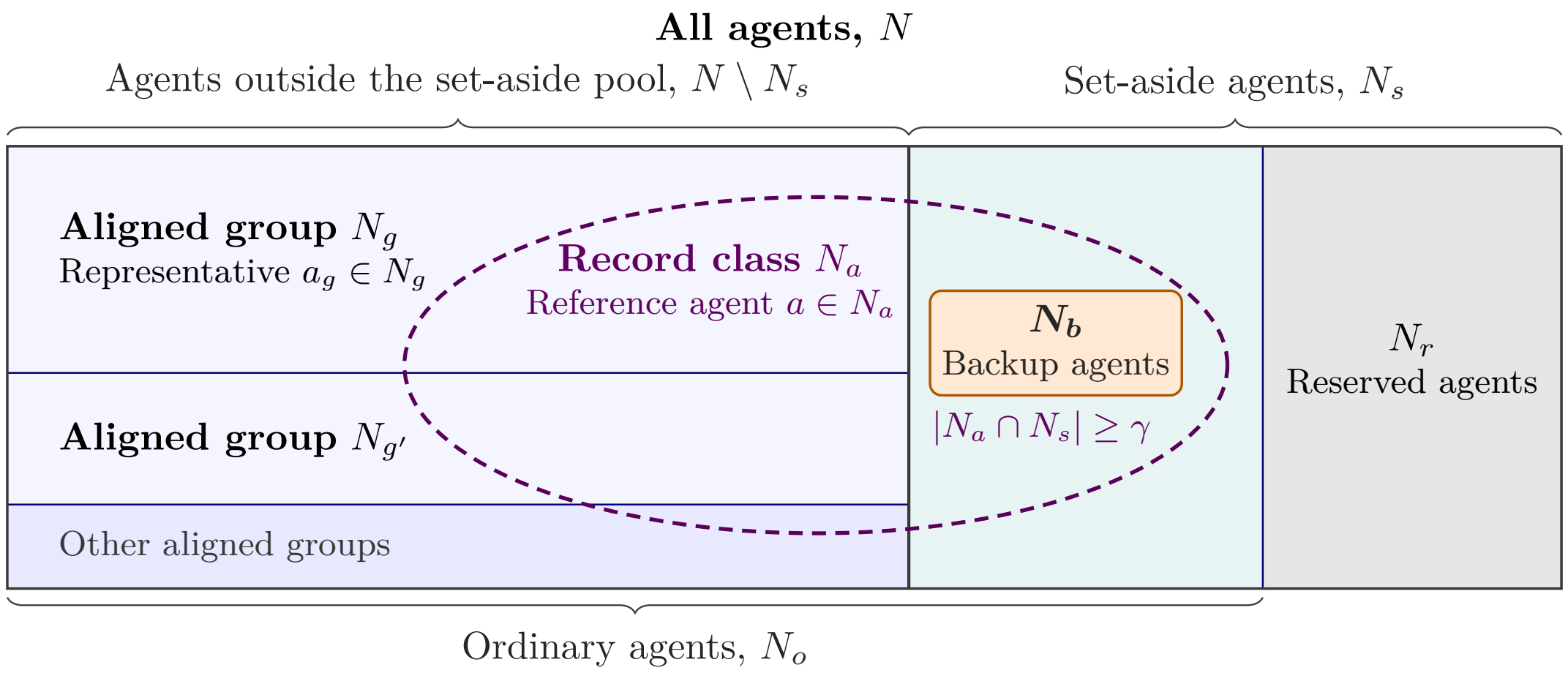

When forming $N_s$, we set aside $\gamma$ members of $N_a$ (recall that $|N_a| = n^{1-o(1)} \gg \gamma$).
Since $k \le \sqrt{\gamma} \le \gamma$, we can assign each of the $k$ backup
bundles to a distinct one of these agents.
Let $N_b$ contain these $k$ \emph{backup agents}.
See Figure~\ref{fig:mms-agent-sets} for an illustration of the agent sets.

On any refined interval $I_\ell$ where $N_a$ has a constant $d_1$-rounded value, a backup agent's requests in $I_\ell$ have no deadline. On each remaining interval $I_\ell$, let $h_\ell$ be the number of backup requests. We pre-assign these requests to local positions $1,3,\dots,2h_\ell-1$. They lie among the first $2\gamma$ positions because $h_\ell\le \kappa\cdot k\le \kappa\cdot\sqrt{\gamma}\le \gamma$, where the choice of $\gamma$ in Appendix~\ref{app:bundle-selection} gives $\sqrt{\gamma}\ge\kappa$.
To share the remaining items $M''_2$ among all set-aside agents $N_s = N_b \cup (N_s\setminus N_b)$, we now fix one small positive rational $\xi \in (0, 1/4)$ satisfying $\kappa\cdot d_1 + {d_2}/{\tau} + \xi + \tau < 1 - \beta$.
This is possible because we chose $\kappa\cdot d_1 + \tau + d_2/\tau < 1-\beta$.

\begin{lemma}\label{lem:exceptional}
In polynomial time, we can divide $M_2$ among $N_s$ so that every backup agent in $N_b$ reaches total value at least $\beta$ (once she receives her prescribed items on $M_1$), and every other set-aside agent $i\in N_s\setminus N_b$ either reaches value at least $\beta$ on $M_2$ alone or satisfies $v_i(M_1) = n - o(n)$.
\end{lemma}

\begin{proof}
Starting from the prescribed assignments on $M_1$, we give each backup
agent the items of her backup bundle that lie in $M'_2$.
We then divide $M''_2$ among all agents in $N_s$.

\medskip\noindent\emph{Step 1: Value of backup agents on $M_1$.}
By Lemma~\ref{lem:representative}(2), each designated odd position gives every backup agent $i\in N_b$ the highest $d_1$-rounded value of $a$ on $I_\ell$. This is at least the rounded value of the item it replaces in the backup bundle $Q$ assigned to $i$. On flat intervals, any item in $I_\ell$ gives the same $d_1$-rounded value as $a$. Since rounding down by $d_1$ loses less than $d_1$ per item across at most $\kappa$ items in $M_1$, each backup agent $i\in N_b$ gets at least $v_a(Q\cap M_1) - \kappa\cdot d_1$.

\medskip\noindent\emph{Step 2: Value of backup agents on $M'_2$.}
Give each backup agent $i\in N_b$ (paired with $Q\in\bP^a$) the exact items $Q\cap M'_2$. These sets are disjoint because all $k$ backup bundles come from $\bP^a$. Since every item in $M'_2$ has value at least $\tau$ to agent $a$ and $v_a(Q) = 1$, we have $|Q\cap M'_2| \le 1/\tau$. Because $i\in N_a$ shares $a$'s $d_2$-rounded values on $M'_2$, her value on $M'_2$ is at least $v_a(Q\cap M'_2) - d_2/\tau$.

\medskip\noindent\emph{Step 3: Sharing $M''_2$ fractionally.}
Assign each agent $i\in N_s$ a fractional share $s_i\in[0,1]$ of $M''_2$:
\begin{itemize}
    \item For each backup agent $i\in N_b$ paired with $Q\in\bP^a$, set $s_i = (1-\xi)\cdot \frac{v_a(Q\cap M''_2)}{v_a(M''_2)}$ if $v_a(M''_2) > 0$, and $s_i = 0$ otherwise. Since the backup bundles are disjoint, $\sum_{i\in N_b} s_i \le 1-\xi$.
    \item If $N_s\setminus N_b$ is nonempty, give each $i\in N_s\setminus N_b$ the equal share $s_i = \xi / |N_s\setminus N_b|$.
\end{itemize}
List $M''_2$ in its common IDO order as local positions $1,\dots,|M''_2|$. For each $i\in N_s$ and each position $b\in[|M''_2|]$, create one request for $i$ with deadline $b$ whenever $\floor{s_i\cdot b} > \floor{s_i\cdot(b-1)}$. Then agent $i$ has $\floor{s_i\cdot b}$ requests due by $b$, and summing over all $i\in N_s$ gives at most $\sum_{i\in N_s} s_i\cdot b \le b$ requests due by $b$. Sorting all requests by deadline and assigning the $j$-th request to the $j$-th item of $M''_2$ meets every deadline, giving agent $i$ a subset $Z_i\subseteq M''_2$ with $|Z_i\cap[b]| \ge \floor{s_i\cdot b} \ge s_i\cdot b - 1$ for every prefix length $b$.
Let $u_1 \ge \dots \ge u_\ell \ge u_{\ell+1} = 0$ be agent $i$'s values for $M''_2$ ($\ell = |M''_2|$). Applying~\eqref{eq:prefixvalue} to $Z_i$ (if $\ell = 0$, she loses nothing) gives
\begin{equation*}
    v_i(Z_i)
    = \sum_{b=1}^{\ell} |Z_i\cap[b]|\cdot(u_b - u_{b+1})
    \ge \sum_{b=1}^{\ell} (s_i\cdot b - 1)\cdot(u_b - u_{b+1})
    = s_i\cdot v_i(M''_2) - u_1.
\end{equation*}
Thus, every $i\in N_s$ receives value at least $s_i\cdot v_i(M''_2)$ minus her single largest item value in $M''_2$.

\medskip\noindent\emph{Step 4: Checking the final values for $N_b$ and $N_s\setminus N_b$.}
For any backup agent $i\in N_b$ (paired with $Q\in\bP^a$), our choice of $a$ gives $v_i(M''_2) \ge v_a(M''_2)$, and every item in $M''_2$ has value at most $u_1 < \tau$ to agent $i$. Using $v_a(Q\cap M''_2) \le v_a(Q) = 1$, her value on $M''_2$ is at least
\begin{equation*}
    (1-\xi)\cdot v_a(Q\cap M''_2) - \tau \ge v_a(Q\cap M''_2) - \xi - \tau.
\end{equation*}
Adding her values on $M_1$, $M'_2$, and $M''_2$ gives total value at least $1 - \kappa\cdot d_1 - \frac{d_2}{\tau} - \xi - \tau > \beta$.
Now take any $i\in N_s\setminus N_b$ whose value from $M_2$ is below $\beta$. Her largest item in $M''_2$ has value $u_1 \le 1$, so her share $s_i = \xi / |N_s\setminus N_b|$ gives
\begin{equation*}
    \frac{\xi}{|N_s\setminus N_b|}\cdot v_i(M''_2) - 1 \le v_i(Z_i) < \beta
    \quad\Longrightarrow\quad
    v_i(M''_2) \le \frac{(\beta + 1)\cdot |N_s\setminus N_b|}{\xi}.
\end{equation*}
Meanwhile, $M'_2$ has at most $T/\tau$ items, each worth at most $1$, so $v_i(M'_2) \le T/\tau$. Since $v_i(M) = n$, $|N_s| = o(n)$, and $T = n^{o(1)}$, we conclude that
\begin{equation*}
    v_i(M_1) \ge n - \left(\frac{(\beta + 1)\cdot |N_s\setminus N_b|}{\xi} + \frac{T}{\tau}\right) = n - o(n).
\end{equation*}
Any leftover unassigned items of $M_2$ are added to agents in $N_s$ without lowering anyone's value. All steps run in polynomial time.
\end{proof}

\subsubsection{Allocation of \texorpdfstring{$M_1$}{}}\label{subsubsec:quota-assignment}

We now combine the $n$ selected bundles from Lemma~\ref{lem:quotas} with the allocation of $M_2$ from Lemma~\ref{lem:exceptional} to allocate $M_1$ so that every agent has value at least $\beta$ (Lemma~\ref{lem:quota-assignment}), completing Theorem~\ref{thm:common-prefix-allocation}.

\begin{lemma}\label{lem:quota-assignment}
We can allocate $M_1$ in polynomial time to meet all backup agents' requirements and give every other agent total value at least $\beta$ across $M_1$ and $M_2$.
\end{lemma}
\begin{proof}
We complete the allocation in two steps: we first fill all requests of the $n$ selected bundles from Lemma~\ref{lem:quotas} with actual items of $M_1$, and then assign the resulting completed bundles.

\smallskip

We first construct the bundles.
In each interval $I_\ell$, we sort the sensitive representative requests by deadline. If the $j$-th request has deadline $b$, at least $j$ requests are due by $b$, so Lemma~\ref{lem:quotas}(2) gives $j<b/4$. We assign this request to local position $2j$. This meets its deadline because $2j\le b$, and all fixed backup positions are odd. We fill the deadline-free requests with the remaining items of $I_\ell$ in any order. The counts agree because the selected bundles request exactly $|I_\ell|$ items by Lemma~\ref{lem:quotas}(1).
We immediately give each backup agent in $N_b$ her completed backup bundle; by Lemma~\ref{lem:exceptional}, her total value across $M_1$ and $M_2$ is at least $\beta$. Next, consider any completed representative bundle of a group $N_g$ and any member $i\in N_g$. Because the bundle is eligible for $a_g$, its original items in $M_1$ have value at least $1 - d_1$ to $a_g$, and there are fewer than $\kappa$ such items. On each interval $I_\ell$ where $a_g$ is insensitive, $i$ and $a_g$ share the same constant $d_1$-rounded value across all of $I_\ell$. On each interval $I_\ell$ where $a_g$ is sensitive, filling the request at or before its checkpoint deadline gives $i$ at least the $d_1$-rounded value that $a_g$ assigned to the original item. Since rounding down loses less than $d_1$ per item, every $i\in N_g$ values every completed representative bundle of $N_g$ at least $1 - d_1 - \kappa\cdot d_1 = 1 - (\kappa+1)\cdot d_1 > \beta$.

\smallskip

Next, we assign the completed representative bundles.
By Lemma~\ref{lem:quotas}(3), each group $N_g$ retains at least $\phi_g\cdot n - \sqrt \gamma$ representative bundles. Using $\vert{}N_g\vert{}\ge\sqrt n$, $\vert{}N_s\vert{}\ge n/L$, and $\gamma = n^{o(1)}$, we have
\begin{equation*}
    \phi_g\cdot n - |N_g| = \phi_g\cdot |N_s| = \frac{|N_g|\cdot |N_s|}{n - |N_s|} \ge \frac{\sqrt n}{L} > \sqrt \gamma.
\end{equation*}
Thus, every group $N_g$ retains at least $\vert{}N_g\vert{}$ completed bundles. Across all groups, there are $n - k$ completed representative bundles for $n - \vert{}N_s\vert{}$ group members, leaving exactly $(n-k) - (n-\vert{}N_s\vert{}) = \vert{}N_s\setminus N_b\vert{}$ extra bundles in total.
We process the agents in $N_s\setminus N_b$ one by one in any order. At each agent's turn, pick any group $N_g$ that currently has strictly more remaining bundles than members, and give her whichever remaining bundle of $N_g$ she values most. 
Once all agents in $N_s\setminus N_b$ have chosen, each group $N_g$ has exactly $\vert{}N_g\vert{}$ bundles left, which we assign arbitrarily to its $\vert{}N_g\vert{}$ members. By Step~1, every group member receives value at least $\beta$.
It remains to check that every agent $i\in N_s\setminus N_b$ who received value less than $\beta$ on $M_2$ in Lemma~\ref{lem:exceptional} chooses a bundle worth at least $\beta$ on $M_1$. Fix the group $N_g$ chosen at her turn, and let $U_g\subseteq M_1$ be the set of all items in $M_1$ belonging to the bundles of $N_g$. For any $j\in[\vert{}M_1\vert{}]$, the prefix $M[j]$ consists of some whole intervals together with a strict prefix of at most one additional interval. By Lemma~\ref{lem:representative}(1), that partial interval contributes at most $\ceil{d_1\cdot \vert{}M_1\vert{}} - 1 \le d_1\cdot \vert{}M_1\vert{} \le \kappa\cdot d_1\cdot n$ items, so the whole intervals inside $M[j]$ contain at least $j - \kappa\cdot d_1\cdot n$ items. By Lemma~\ref{lem:quotas}(3), the total shortfall of $U_g$ below $\phi_g\cdot \vert{}I_\ell\vert{}$ across all intervals $I_\ell$ is at most $\kappa\cdot\sqrt \gamma$. Counting only the items of $U_g$ in the whole intervals contained in $M[j]$ therefore gives
\begin{equation*}
|U_g\cap M[j]| \ge \phi_g\cdot\left(j - \kappa\cdot d_1\cdot n\right) - \kappa\cdot\sqrt \gamma
\qquad \text{for all } j\in[|M_1|].
\end{equation*}
Then we can convert this prefix count bound into a value bound (via a telescoping sum):
\begin{equation*}
v_i(U_g) \ge \phi_g\cdot\left(v_i(M_1) - \kappa\cdot d_1\cdot n\right) - \kappa\cdot\sqrt \gamma.
\end{equation*}

By Lemma~\ref{lem:quotas}(3), group $N_g$ started Step~2 with at most $\phi_g\cdot n$ bundles, so earlier agents in $N_s\setminus N_b$ took at most $\phi_g\cdot n - \vert{}N_g\vert{} = \phi_g\cdot \vert{}N_s\vert{}$ bundles from $N_g$. Each bundle contains fewer than $\kappa$ items from $M_1$, each worth at most $1$ to agent $i$, so earlier agents removed at most $\kappa\cdot\phi_g\cdot \vert{}N_s\vert{}$ of value from $U_g$. Dividing the remaining value by $\phi_g\cdot n$ (an upper bound on the number of remaining bundles in $N_g$) shows that agent $i$'s favorite remaining bundle has value at least
\begin{equation*}
\frac{v_i(M_1)}{n} - \kappa\cdot d_1 - \frac{\kappa\cdot\sqrt \gamma}{\phi_g\cdot n} - \frac{\kappa\cdot |N_s|}{n}
\geq 1 - \kappa\cdot d_1 - o(1) > \beta,
\end{equation*}
where the first inequality follows since $v_i(M_1) = n - o(n)$ (by Lemma~\ref{lem:exceptional}), $\phi_g\cdot n \ge \vert{}N_g\vert{} \ge \sqrt n$, $\gamma = n^{o(1)}$, and $\vert{}N_s\vert{} = o(n)$.
Thus, every agent reaches total value at least $\beta$. Because Steps~1 and~2 only sort requests and compare completed bundles, the procedure runs in polynomial time.
\end{proof}

\begin{theorem}\label{thm:common-prefix-allocation}
For every fixed rational $\beta\in(0,1)$ and sufficiently large $n$, we can compute a $\beta$-MMS allocation in polynomial time for a normalized IDO instance with $m=O(n)$ goods and supplied MMS partitions that satisfies Properties~(\ref{prop:reserved-small})--(\ref{prop:prefix-count}).
\end{theorem}
\begin{proof}
Apply Lemma~\ref{lem:representative} (with $\gamma$ chosen as in Appendix~\ref{app:bundle-selection}) to obtain the intervals $I_1,\dots,I_q$, set-aside pool $N_s$, and aligned groups $\{N_g\}$. Lemma~\ref{lem:quotas} then selects $n-k$ eligible group bundles and $k\le \sqrt \gamma$ backup bundles from $\bP^a$ matching every $|I_\ell|$ with $< b/4$ sensitive load, Lemma~\ref{lem:exceptional} divides $M_2$ among $N_s$ and pre-assigns the backup agents' positions, and Lemma~\ref{lem:quota-assignment} completes the allocation of $M_1$ so that every agent receives value at least $\beta$. Any leftover items in $M_2$ are assigned arbitrarily, and every stage runs in polynomial time.
\end{proof}

\subsection{Putting Everything Together}\label{subsec:putting-together}

We now combine the reduction and LP of Section~\ref{subsec:lp-framework} with the large-slack and small-slack algorithms of Sections~\ref{subsec:slack-rounding}--\ref{subsec:completing-allocation} to prove Theorem~\ref{thm:cardinal-goods}.

\begin{proof}[Proof of Theorem~\ref{thm:cardinal-goods}]
We fix a rational $\beta\in(1-\varepsilon/2,1-\varepsilon/4)$. By Lemma~\ref{lem:reduction} with $\alpha=1-\varepsilon$ and $\beta$, we construct in polynomial time a normalized IDO instance with $m=O(n)$ goods and MMS partitions $\{\bP^i\}_{i\in N}$ so that any $\beta$-MMS allocation maps back to a $(1-\varepsilon)$-MMS allocation of the original instance in polynomial time. We fix a rational $\sigma\in(\beta,1)$ and solve the LP of Section~\ref{subsec:lp-framework} in polynomial time to obtain a basic optimum $(\lambda,x)$ and its optimal dual prices.

When $\Delta\ge2R$, Algorithm~\ref{alg:slack} and Lemma~\ref{lem:slack-rounding} compute in polynomial time an allocation giving every agent value at least $\sigma>\beta$. 
When $\Delta<2R$, Section~\ref{subsec:valuable-prefix} gives Properties~(\ref{prop:reserved-small})--(\ref{prop:prefix-count}), so we apply Theorem~\ref{thm:common-prefix-allocation} to compute a $\beta$-MMS allocation.
In either case, mapping the resulting $\beta$-MMS allocation back produces a $(1-\varepsilon)$-MMS allocation of the original instance in polynomial time.
\end{proof}


\section*{Declaration for the Use of AI}
We used OpenAI's GPT-6.2 to assist with mathematical analysis, language editing, and LaTeX formatting. All mathematical statements, proofs, calculations, and citations were reviewed and verified by the authors, who take full responsibility for the content of the paper.

\bibliographystyle{alpha}
\bibliography{MMS}

\addtocontents{toc}{\protect\setcounter{tocdepth}{1}}
\appendix

\section{Missing Analysis from Section~\ref{sec:dichotomy}}\label{app:proofs}

\subsection{Normalization and Reduction of Goods}\label{app:compression}
\begin{figure}[htbp]
  \centering
  \begin{tikzpicture}[
    x=1cm,y=1cm,font=\small,
    item/.style={draw=black!25,fill=black!3,rounded corners=2pt,
      minimum width=6.6mm,minimum height=5.5mm,inner sep=0pt},
    leading/.style={item,draw=cyan!50!black,fill=cyan!12},
    trailing/.style={item,draw=orange!65!black,fill=orange!14}
  ]
    \node[font=\small\bfseries,anchor=west] at (0,2.58) {First $h$ items};
    \foreach \j in {1,...,4}{
      \node[item] at ({0.36+0.77*(\j-1)},1.38) {$e_{\j}$};
    }
    \node[text=black!65,anchor=west,font=\footnotesize] at (0,0.48)
      {Keep as single goods.};
    \draw[black!15] (3.2,0.08) -- (3.2,2.62);
    \node[font=\small\bfseries,anchor=west] at (3.64,2.58)
      {Deal the remaining items cyclically into $h$ columns};
    \foreach \x/\j/\k/\ell/\style in {
      4.02/5/9/13/leading,
      4.86/6/10/14/item,
      5.70/7/11/15/item,
      6.54/8/12/16/trailing}{
      \node[\style] at (\x,1.92) {$e_{\j}$};
      \node[\style] at (\x,1.10) {$e_{\k}$};
      \node[\style] at (\x,0.28) {$e_{\ell}$};
    }
    \draw[-{Stealth[length=3pt]},cyan!55!black,thick] (4.02,2.30)
      -- (6.54,2.30);
    \draw[-{Stealth[length=3pt]},orange!70!black,thick,rounded corners=1pt]
      (6.54,1.58) -- (6.54,1.50) -- (4.02,1.50) -- (4.02,1.42);
    \node[anchor=west,text=black!72,font=\footnotesize]
      at (7.22,1.62) {Each column becomes one good.};
    \node[anchor=west,align=left,text=black!72,font=\footnotesize]
      at (7.22,0.98)
      {The first item of each round follows\\
       the last item of the previous round.};
  \end{tikzpicture}
  \caption{Cyclic compression, illustrated with $h=4$. In every ordered row, the first and last columns interleave in the common order, so their totals differ by at most the value of $e_{h+1}$.}
  \label{fig:cyclic-compression}
\end{figure}
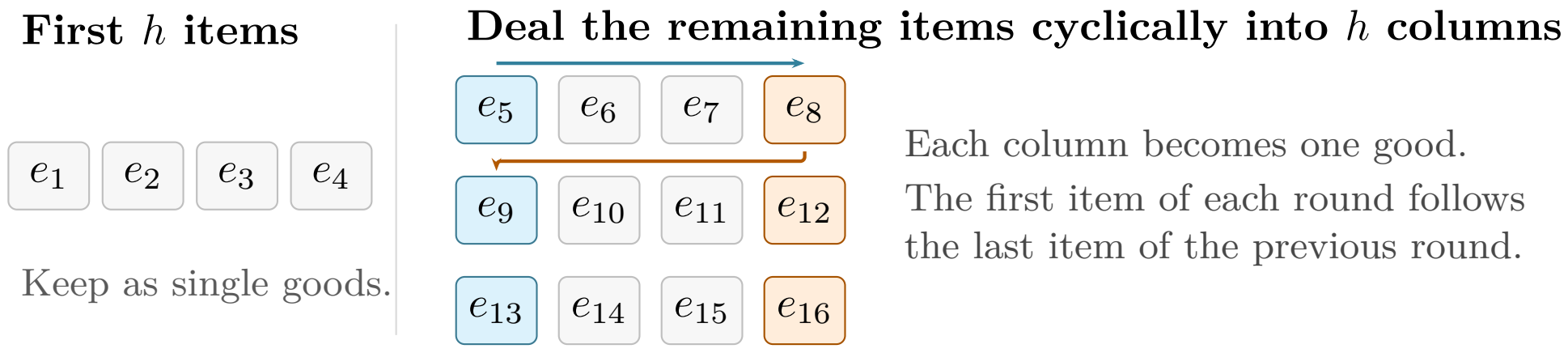

In this subsection, we prove Lemma~\ref{lem:reduction}, which reduces any goods instance to a normalized IDO instance with $O(n)$ goods and explicit MMS partitions.
We first isolate this compression step in Lemma~\ref{lem:compress} (illustrated in Figure~\ref{fig:cyclic-compression}), which merges $M$ into at most $2h$ items while constructing new partitions whose corresponding bundle values differ from the original ones by at most $2n/h$.

\begin{lemma}\label{lem:compress}
Given an IDO instance with $v_i(M)\le n$ for all $i\in N$,
an $n$-partition $\bP^i$ of $M$ for each $i\in N$, and an integer $h\ge n$,
we can in polynomial time merge $M$ into at most $2h$ items and compute
$n$-partitions $\tilde\bP^i$ of these items satisfying
$|v_i(\tilde P_j^i)-v_i(P_j^i)|\le 2n/h$ for all $i\in N$ and $j\in[n]$.
\end{lemma}
\begin{proof}
If $m\le2h$, we keep each item as a singleton and set $\tilde\bP^i=\bP^i$. Otherwise, we keep $e_1,\dots,e_h$ as singletons and deal $M\setminus M[h]$ cyclically into $h$ columns $C_1,\dots,C_h$, where $C_k=\{e_{th+k}: t\ge 1,\ th+k\le m\}$ (Figure~\ref{fig:cyclic-compression}). Each column $C_k$ becomes one compressed item of value $v_i(C_k)=\sum_{e\in C_k}v_i(e)$, giving $2h$ compressed items in total. 
For any $i\in N$, since $v_i(M)\le n$, we have $v_i(e_{h+1})\le n/h$ and $v_i(C_1)\ge\dots\ge v_i(C_h)$, which yields $v_i(C_h)\le n/h$ as $\sum_{k=1}^h v_i(C_k)\le n$. 
Furthermore, by construction, we have $v_i(C_1\setminus\{e_{h+1}\})\le v_i(C_h)$, which implies that 
$$v_i(C_k)\le v_i(C_1)\le v_i(e_{h+1})+v_i(C_h)\le 2n/h \qquad\text{ for all } k\in[h]. $$

For each agent $i\in N$, we initialize $\tilde P_j^i=P_j^i\cap M[h]$ for every $j\in[n]$ and track each bundle's deficit $\delta_j=v_i(P_j^i)-v_i(\tilde P_j^i)$. Initially $\delta_j=v_i(P_j^i\setminus M[h])\ge0$, and $\sum_{j=1}^n\delta_j=v_i(M\setminus M[h])=\sum_{k=1}^h v_i(C_k)$. We assign the columns $C_1,\dots,C_h$ one by one, each time adding $C_k$ to a bundle $j$ with maximum current deficit $\delta_j$ (decreasing $\delta_j$ by $v_i(C_k)$). 
Before each step, $\sum_{j=1}^n\delta_j$ equals the total value of the unassigned columns and is therefore nonnegative, so the chosen bundle $j$ has $\delta_j\ge0$ before receiving $C_k$ and $\delta_j\ge-2n/h$ afterward. Since all deficits start nonnegative, every bundle satisfies $\delta_j\ge-2n/h$. 
On the other hand, we show that every bundle also satisfies $\delta_j\le 2n/h$ at termination. If some bundle $j$ ended with $\delta_j>2n/h$, then because deficits never increase, $\delta_j>2n/h$ would hold before every step. Consequently, at every step the chosen bundle would have deficit above $2n/h$ before receiving a column, so its deficit would remain positive afterward. 
However, every bundle would therefore end with nonnegative deficit while $\delta_j>0$, contradicting $\sum_{j'=1}^n\delta_{j'}=0$.
Thus, $|\delta_j|=|v_i(P_j^i)-v_i(\tilde P_j^i)|\le2n/h$ for all $j\in[n]$, and the procedure runs in polynomial time.
\end{proof}

\begin{proof}[Proof of Lemma~\ref{lem:reduction}]
We choose fixed rational parameters $\beta_0\in(\alpha,(\alpha+\beta)/2)$ and $\varepsilon_1\in(0,1)$ such that $(1-\varepsilon_1)\cdot\beta_0\ge\alpha$.
In the following three steps, we construct the reduced instance and each agent's MMS partition in the reduced instance.

\smallskip

\noindent\textbf{Initial Normalization.}
For each agent $i\in N$, we apply the machine covering scheme~\cite{journals/orl/Woeginger97} with item sizes $v_i(e)$ on $n$ identical machines to compute a partition $\bP^i=(P_1^i,\dots,P_n^i)\in\pi_n(M)$ whose minimum bundle value $v_i^{\min}=\min_{j\in[n]}\{v_i(P_j^i)\}$ satisfies $v_i^{\min}\ge(1-\varepsilon_1)\cdot\mu_i$. When $v_i^{\min}>0$, we apply the bundlewise rescaling of Akrami, Garg, Sharma, and Taki~\cite{akrami2023simplification} and set $\hat v_i(e)=v_i(e)/v_i(P_j^i)$ for every $j\in[n]$ and $e\in P_j^i$. Since $v_i(P_j^i)\ge v_i^{\min}$ for every bundle $P_j^i$, this gives $\hat v_i(P_j^i)=1$ and $v_i(S)\ge v_i^{\min}\cdot\hat v_i(S)$ for all $S\subseteq M$. When $v_i^{\min}=0$, it follows that $\mu_i=0$, so we replace $\bP^i$ with an arbitrary partition of $M$ into $n$ nonempty bundles (using $m\ge n$) and define $\hat v_i$ by assigning value $1$ to one chosen item in each bundle and $0$ to all other items. The resulting instance $\mathcal I_1=(N,M,\hat\bv)$ is normalized with MMS partitions $\{\bP^i\}_{i\in N}$, and every $\beta_0$-MMS allocation $\bX$ of $\mathcal I_1$ satisfies $v_i(X_i)\ge v_i^{\min}\cdot\hat v_i(X_i)\ge(1-\varepsilon_1)\cdot\beta_0\cdot\mu_i\ge\alpha\cdot\mu_i$ for all $i\in N$.

\smallskip

\noindent\textbf{Compression, Renormalization, and Final IDO Reduction.}
We first apply Lemma~\ref{lem:ido} to $\mathcal I_1$ by sorting each agent's valuation row in nonincreasing order and applying the same permutation to $\bP^i$, which keeps every bundle value equal to $1$. We set $h=\ceil{2\beta\cdot n/(\beta-\beta_0)}$, so that $h>2n$ and $(1-2n/h)\cdot\beta\ge\beta_0$.
By Lemma~\ref{lem:compress}, we obtain a compressed instance with at most $2h=O(n)$ items (with implicit constant $O(1/(\beta-\alpha))$) and $n$-partitions $\{\tilde\bP^i\}_{i\in N}$ whose bundle values all lie in $[1-2n/h,1+2n/h]$.
For each agent $i\in N$ and each bundle $\tilde P_j^i\in\tilde\bP^i$, we divide the value of every compressed item in $\tilde P_j^i$ by the bundle's value (which is at least $1-2n/h>0$). This yields a normalized instance on the compressed items with MMS partitions $\{\tilde\bP^i\}_{i\in N}$. Because both merging items and rescaling bundles can disrupt the common item ordering, this compressed normalized instance need not be IDO. We therefore apply Lemma~\ref{lem:ido} one final time, sorting each agent's compressed items in nonincreasing order and applying the same permutation to $\tilde\bP^i$. Each permutation preserves unit bundle values, yielding a normalized IDO instance $\mathcal I_2$ with at most $2h=O(n)$ goods and explicit MMS partitions.

\smallskip

\noindent\textbf{Mapping Allocations Back.}
Given any $\beta$-MMS allocation of $\mathcal I_2$, Lemma~\ref{lem:ido} maps it back to an allocation of the compressed normalized instance in which every agent has value at least $\beta$. Every bundle of $\tilde\bP^i$ had value at least $1-2n/h$ before rescaling. Thus, each compressed item's value before rescaling is at least $1-2n/h$ times its normalized value to agent $i$, and summing over the allocated items gives each agent value at least $(1-2n/h)\cdot\beta\ge\beta_0$ before rescaling. Each compressed item then expands into its constituent items without changing any bundle value, and Lemma~\ref{lem:ido} maps the resulting allocation back to a $\beta_0$-MMS allocation of $\mathcal I_1$. By the initial normalization guarantee, this allocation is $\alpha$-MMS for the original instance $\mathcal I$. Both the construction and the reverse mapping run in polynomial time.
\end{proof}

\subsection{Proof of Lemma~\ref{lem:quotas}}\label{app:bundle-selection}

\begin{proof}[Proof of Lemma~\ref{lem:quotas}]
We build the selection in two stages: first, round the eligible representative bundles at a vertex of an LP (and drop fractional and ineligible bundles), then repair the exact interval counts using at most $\sqrt \gamma$ bundles from $\bP^a$. At the end of Stage~2, we read off the exact formula for $\gamma$ needed to absorb all rounding and repair errors.

\smallskip\noindent\emph{Stage 1: Vertex rounding of eligible representative bundles.}
Introduce one variable $z_{g,j}\in[0,1]$ for each bundle $P_j^{a_g}$ of each representative $a_g$, initialized at $z_{g,j} = \phi_g$. Initially, group $N_g$ has total bundle weight $\phi_g\cdot n$ and total request weight $\phi_g\cdot |I_\ell|$ in each interval $I_\ell$. Across all groups, the total bundle weight is $n$ and the total request weight in $I_\ell$ is $|I_\ell|$.

First, discard all ineligible bundles (set $z_{g,j} = 0$ whenever $P_j^{a_g}$ is ineligible). How much weight does this discard? Because $v_{a_g}(M_2) \le T$ and each ineligible bundle has value strictly greater than $d_1$ on $M_2$, each representative $a_g$ has fewer than $T/d_1$ ineligible bundles. Summing over all groups using $\sum_g \phi_g = 1$, the total weight discarded here is at most $\sum_g \phi_g\cdot(T/d_1) = T/d_1$.

Next, consider the polytope of all weight vectors $z\in[0,1]$ on the remaining \emph{eligible} bundles that preserve the following linear equalities (keeping every left-hand side equal to its current value after discarding ineligible bundles):
\begin{itemize}
    \item For each group $N_g$: preserve the group's total bundle weight ($1$ equality) and its total request weight in each interval $I_\ell$ ($q$ equalities). Across at most $G$ groups, this is at most $G\cdot(q+1)$ equalities.
    \item For each interval $I_\ell$: preserve the total fractional number of sensitive requests due by local position $b'$ at every power-of-two checkpoint $b' \in \{1, 2, 4, 8, \dots\}\cap[|I_\ell|]$ and at the last position $b' = |I_\ell|$. Since $|I_\ell| \le m$, this adds at most $2 + \ceil{\log_2 m}$ checkpoints per interval, so at most $q\cdot(2+\ceil{\log_2 m})$ equalities in total.
\end{itemize}
In total, this polytope is defined by $0 \le z_{g,j} \le 1$ together with at most $G\cdot(q+1) + q\cdot(2+\ceil{\log_2 m})$ linear equalities.

Pick a vertex $z^*$ of this polytope. How many coordinates of $z^*$ can be strictly fractional ($0 < z^*_{g,j} < 1$)? At most the number of linear equalities! Indeed, if there were strictly more fractional coordinates than linear equalities, then the homogeneous linear system formed by those equalities restricted to the fractional coordinates would have a nonzero solution $v$. Moving from $z^*$ to $z^* + \eta v$ and $z^* - \eta v$ for a tiny $\eta > 0$ would stay feasible, contradicting that $z^*$ is a vertex. Moreover, we can compute a vertex in polynomial time by looping over the coordinates one by one: at each step, minimize the current coordinate over the current feasible face (solving one LP) and fix that coordinate to its minimum value before moving to the next coordinate.

Now let us check that $z^*$ obeys the $< b/4$ sensitive deadline bound in every interval $I_\ell$ and at every position $b\in[|I_\ell|]$. Let $b'$ be the last checkpoint in $I_\ell$ at or before $b$. Since sensitive requests are due only at checkpoints, the load at $b$ equals the load at $b'$. Look at the initial (pre-rounding) sensitive load at $b'$:
\begin{itemize}
    \item For a sensitive request with original local position $r$ and deadline $c$, we have $r<2c$. If $c$ is the last position of $I_\ell$, then $r=c$. Otherwise, $c$ is a power of two, and $r\ge2c$ would make $2c$ a later checkpoint at or before $r$. Hence, a request due by $b'$ comes from a position $r<2c\le2b'\le2b$. Thus, each sensitive representative's partition $\bP^{a_g}$ contributes at most $2b$ sensitive requests due by $b'$.
    \item If a representative $a_g$ is sensitive to $I_\ell$, then every member of $N_g$ is also sensitive to $I_\ell$ (since they share the two different rounded values at the first and last checkpoints of $I_\ell$). Because at most $n/10 + 1$ ordinary agents are sensitive to $I_\ell$, the sum of $\phi_g = |N_g|/(n-|N_s|)$ over all sensitive groups is at most $(n/10 + 1)/(n - |N_s|)$.
\end{itemize}
Because discarding ineligible bundles only lowered the load at $b'$ and moving to $z^*$ preserved it, the sensitive load of $z^*$ at position $b$ equals its load at $b'$, which is at most $2b\cdot \frac{n/10 + 1}{n - |N_s|} < \frac{b}{4}$ for sufficiently large $n$ (since $|N_s| = o(n)$).

Now discard all fractional variables of $z^*$ (setting every $z^*_{g,j}\in(0,1)$ to $0$). Counting both the ineligible weight ($\le T/d_1$) and the fractional variables (at most $G\cdot(q+1) + q\cdot(2+\ceil{\log_2 m})$ of them, each of weight $< 1$), the total bundle weight discarded across Stage~1 is at most
\begin{equation*}
    J = \frac{T}{d_1} + G\cdot(q+1) + q\cdot\left(2 + \ceil{\log_2 m}\right).
\end{equation*}
The surviving variables all equal $1$, giving a collection of distinct eligible representative bundles whose sensitive deadline load is still strictly below $b/4$ everywhere.

\smallskip\noindent\emph{Stage 2: Exact inventory repair using bundles from $\bP^a$.}
For any bundle $P$, call its vector of interval counts $(|P\cap I_1|,\dots,|P\cap I_q|)\in\mathbb Z_{\ge 0}^q$ its \emph{configuration}. Let $\mathcal C$ be the set of all nonnegative integer vectors of length $q$ whose coordinates sum to at most $\kappa$. Adding one extra coordinate equal to $\kappa$ minus the sum gives nonnegative integer vectors of length $q+1$ summing to $\kappa$, so by stars and bars, $|\mathcal C| = \binom{q + \kappa}{\kappa} = O(1)$.
For each configuration $c\in\mathcal C$:
\begin{itemize}
    \item let $x_c$ be the number of surviving eligible representative bundles of configuration $c$ (summed across all groups), and
    \item let $h_c$ be the number of bundles in the reference partition $\bP^a$ of configuration $c$.
\end{itemize}
Form the $(q+1)\times|\mathcal C|$ matrix $A$ whose column for configuration $c\in\mathcal C$ is the vector $(1, c_1, \dots, c_q)^T$. The first row of $A$ counts bundles, and rows $2,\dots,q+1$ count items in $I_1,\dots,I_q$. Because $\bP^a$ is a complete partition of $M_1$ into $n$ bundles, $A\cdot h = (n, |I_1|, \dots, |I_q|)^T$. Meanwhile, our initial full fractional collection also had totals $A\cdot h$, and Stage~1 only discarded total bundle weight at most $J$ (each unit of bundle weight carrying at most $\kappa$ items in $M_1$). Therefore, the shortage vector $b = A\cdot(h - x)$ is nonnegative in every coordinate and satisfies $\|b\|_1 \le (\kappa + 1)\cdot J$.
Set $H = (\kappa + 2)^{q+2} = O(1)$. We claim that there exists an integer vector $y\in\mathbb Z_{\ge 0}^{|\mathcal C|}$ lying coordinatewise between $x$ and $h$ (meaning $\min\{x_c, h_c\} \le y_c \le \max\{x_c, h_c\}$ for every $c\in\mathcal C$) such that
\begin{equation*}
    A\cdot y = A\cdot h
    \qquad\text{and}\qquad
    \|y - x\|_1 \le H\cdot\left(\|b\|_1 + |\mathcal C| + q + 1\right).
\end{equation*}
We prove this claim in four elementary steps:
\begin{enumerate}
    \item \textbf{Write the difference as a nonnegative nullspace vector.} Define a sign $s_c \in \{+1, -1\}$ for each $c\in\mathcal C$ matching the sign of $h_c - x_c$ (choosing $+1$ if $h_c = x_c$), and let $A_{\text{signed}}$ be the matrix obtained by multiplying column $c$ of $A$ by $s_c$. Then the nonnegative vector $u = (|h - x|, b)\in\mathbb R_{\ge 0}^{|\mathcal C|+q+1}$ satisfies $[A_{\text{signed}}, -I]\cdot u = A(h-x) - b = 0$. If $u = 0$, we simply set $y = x$ and are done.
    \item \textbf{Decompose $u$ into basic vertices (Carath\'eodory's theorem).} If $u \ne 0$, divide $u$ by its coordinate sum $\|u\|_1$ to get a point in the polytope $\{w \ge 0 : [A_{\text{signed}}, -I]\cdot w = 0,\ \sum_r w_r = 1\}$. Because this polytope lies in a hyperplane of dimension $|\mathcal C| + q$, Carath\'eodory's theorem expresses $u/\|u\|_1$ as a convex combination $\sum_{r=1}^t \lambda_r v^{(r)}$ of at most $t \le |\mathcal C| + q + 1$ vertices of this polytope.
    \item \textbf{Clear denominators using Cramer's rule and Hadamard's bound.} Take any vertex $v^{(r)}$. How many positive coordinates can $v^{(r)}$ have? At most $q + 2$: because the polytope has $q + 1$ nullspace equalities plus $1$ sum-to-one equality ($q + 2$ equalities in total), if $v^{(r)}$ had more than $q + 2$ positive coordinates, we could move slightly in both directions along a nonzero vector supported on those coordinates while preserving all $q + 2$ equalities, contradicting that $v^{(r)}$ is a vertex. By the exact same reasoning, the columns of the augmented matrix (the $q+1$ nullspace rows plus the all-ones sum row) on the support of $v^{(r)}$ are linearly independent, so we can extract a nonsingular square submatrix of size at most $(q+2)\times(q+2)$ whose unique solution gives the nonzero entries of $v^{(r)}$. By Cramer's rule, its absolute determinant $D_r$ is a positive integer common denominator for $v^{(r)}$, so $w^{(r)} = D_r v^{(r)}$ is a nonnegative \emph{integer} vector in the nullspace of $[A_{\text{signed}}, -I]$. Moreover, every column of this submatrix has at most one entry of magnitude $1$ in the first row, entries whose absolute values sum to at most $\kappa$ in the next $q$ rows (or a single $-1$), and one entry $1$ in the sum row, so its Euclidean norm is at most $\kappa + 2$. By Hadamard's inequality, $D_r \le (\kappa + 2)^{q+2} = H$. Since the coordinates of $v^{(r)}$ sum to $1$, the integer vector $w^{(r)}$ has coordinate sum $\|w^{(r)}\|_1 = D_r \le H$.
    \item \textbf{Drop zero-residual pieces (zero-sum swaps).} Rescaling $u/\|u\|_1 = \sum_{r=1}^t \lambda_r v^{(r)}$ back to $u$ gives $u = \sum_{r=1}^t \gamma_r w^{(r)}$ with $\gamma_r \ge 0$. Split each coefficient into its integer and fractional parts, $\gamma_r = \floor{\gamma_r} + (\gamma_r - \floor{\gamma_r})$, and set
    \begin{equation*}
        u_{\text{rem}} = \sum_{r=1}^t \left(\gamma_r - \floor{\gamma_r}\right) w^{(r)} = u - \sum_{r=1}^t \floor{\gamma_r}\, w^{(r)}.
    \end{equation*}
    Because both $u$ and every $\floor{\gamma_r} w^{(r)}$ are integer vectors, $u_{\text{rem}}$ is a nonnegative integer nullspace vector, and since $\gamma_r - \floor{\gamma_r} < 1$ and $\|w^{(r)}\|_1 \le H$, its coordinate sum is at most $H\cdot t \le H\cdot(|\mathcal C|+q+1)$. Now form $u'$ by keeping $u_{\text{rem}}$ together with \emph{only} those floored terms $\floor{\gamma_r} w^{(r)}$ whose vector $w^{(r)}$ has at least one nonzero entry in its last $q+1$ coordinates (dropping every term whose last $q+1$ coordinates are all zero---those are zero-sum swaps). Then:
    \begin{itemize}
        \item Clearly $0 \le u' \le u$ coordinatewise, and dropping terms with zero last $q+1$ coordinates leaves the last $q+1$ coordinates unchanged, so the last $q+1$ coordinates of $u'$ still equal $b$.
        \item Every kept floored term $r$ contributes at least $\floor{\gamma_r}$ to the sum of the last $q+1$ coordinates of $u$, which is $\|b\|_1$. Since all terms are nonnegative, $\sum_{r \text{ kept}} \floor{\gamma_r} \le \|b\|_1$.
        \item Therefore, the total coordinate sum of $u'$ is at most
        \begin{equation*}
            \|u'\|_1 \le \|u_{\text{rem}}\|_1 + H\cdot \sum_{r \text{ kept}} \floor{\gamma_r} \le H\cdot\left(\|b\|_1 + |\mathcal C| + q + 1\right).
        \end{equation*}
    \end{itemize}
    Finally, restore the original signs $s_c$ to the first $|\mathcal C|$ coordinates of $u'$ and add them to $x$ to form $y$. Because $0 \le u' \le u$, $y$ is an integer vector lying coordinatewise between $x$ and $h$, satisfies $A(y - x) = b$ (hence $A\cdot y = A\cdot h$), and satisfies $\|y - x\|_1 \le \|u'\|_1 \le H\cdot(\|b\|_1 + |\mathcal C| + q + 1)$, proving the claim.
\end{enumerate}

To find $y$ in polynomial time, we do not need to construct the decomposition above: because $|\mathcal C| = O(1)$ is a constant and every coordinate $y_c$ is an integer in $\{0, 1, \dots, n\}$, we simply loop over all at most $(n+1)^{|\mathcal C|}$ integer vectors $y$ between $x$ and $h$ that satisfy $A\cdot y = A\cdot h$ and pick one minimizing $\|y - x\|_1$.

We now execute the repair vector $y$:
\begin{itemize}
    \item For each configuration $c$ with $y_c < x_c$, discard $x_c - y_c$ more eligible representative bundles of configuration $c$ (from any groups holding them).
    \item For each configuration $c$ with $y_c > x_c$, select $y_c - x_c$ distinct backup bundles of configuration $c$ from $\bP^a$. Enough bundles exist in $\bP^a$ because $y_c \le h_c$.
\end{itemize}
Let $k = \sum_{c:\, y_c > x_c} (y_c - x_c)$ be the total number of backup bundles selected. Since $A\cdot y = A\cdot h$, our $n-k$ remaining eligible representative bundles together with our $k$ backup bundles have total count $n$ and match every interval inventory $|I_\ell|$ exactly.

Finally, we fix the scale $\gamma$ so that it absorbs both Stage~1 and Stage~2. Write $q_{\max} = \floor{12/d_1} + 1$ (so $q \le q_{\max}$), $G_{\max} = \bigl(2+\ceil{\log_2(\kappa n)}\bigr)^{q_{\max}(1+\floor{1/d_1})}$ (so $G \le G_{\max}$), $J_{\max} = T/d_1 + G_{\max}\cdot(q_{\max}+1) + q_{\max}\cdot(2+\ceil{\log_2 m})$ (so $J \le J_{\max}$), and $H_{\max} = (\kappa+2)^{q_{\max}+2}$ (so $H \le H_{\max}$). Notice that every one of these upper bounds depends only on $d_1, \kappa, T, m, n$ and is computable in polynomial time \emph{before} running Lemma~\ref{lem:representative}. Setting
\begin{equation*}
    \sqrt \gamma = \left\lceil 2L + J_{\max} + H_{\max}\cdot\left((\kappa+1)\cdot J_{\max} + \binom{q_{\max}+\kappa}{\kappa} + q_{\max} + 1\right)\right\rceil
\end{equation*}
gives $\gamma = n^{o(1)}$ and guarantees
\begin{equation*}
    J + \|y - x\|_1 \le J + H\cdot\left((\kappa+1)\cdot J + |\mathcal C| + q + 1\right) \le \sqrt \gamma.
\end{equation*}
In particular, $k \le \|y - x\|_1 \le \sqrt \gamma$. In Stage~1, moving to the vertex preserved every group's eligible bundle count and interval request counts; every other step only discarded representative weight. Thus, each group $N_g$ loses at most $\sqrt \gamma$ bundles in total (staying between $\phi_g\cdot n - \sqrt \gamma$ and $\phi_g\cdot n$), retains at most $\phi_g\cdot |I_\ell|$ requests in every $I_\ell$, and has total request shortfall at most $\kappa\cdot\sqrt \gamma$ across all intervals (since each discarded bundle carries fewer than $\kappa$ items in $M_1$). Discarding bundles in Stage~2 only decreases sensitive requests, so the $< b/4$ bound still holds everywhere.
\end{proof}

\section{Implementation using Bundle Comparisons}
\label{sec:comparisons}

The algorithm in Section~\ref{sec:dichotomy} takes the agents' item values as input. In the comparison-query model~\cite{conf/wine/BuLLST24}, each agent is asked to choose between two bundles of goods. Her answer reveals only which bundle she prefers, so the numerical values used in the normalization step are unavailable. In this model, $O(n^4\log m)$ bundle comparisons suffice to compute a $1/2$-MMS allocation. 

In this section, we obtain the $(1-\varepsilon)$-MMS guarantee of Theorem~\ref{thm:cardinal-goods} for every fixed $\varepsilon\in(0,1)$ and sufficiently large $n$ using polynomially many bundle comparisons of the following form.

\begin{definition}\label{def:comparison}
Given an agent $i\in N$ and two bundles $X,Y\subseteq M$, the query $\Comp_i(X,Y)$ returns a bundle of larger value between $X$ and $Y$ according to $v_i$, breaking ties arbitrarily on each query.
\end{definition}

To apply the algorithm of Section~\ref{sec:dichotomy} in this model, we construct an \emph{auxiliary additive valuation} $w_i$ for each agent $i$. We require that any bundle meeting the approximate MMS target under $w_i$ also gives agent $i$ the desired guarantee under her true valuation. Once these auxiliary valuations are known, we can run the algorithm on them without further comparison queries.
Specifically, we use bundle comparisons in two steps. First, we partition the goods into $O(n^2/\varepsilon)$ groups while preserving most of each agent's MMS (Lemma~\ref{lem:rank-groups}). Each auxiliary valuation can then be specified by one value per group. Next, we compare unions of groups to construct the auxiliary valuations with the required guarantee (Lemma~\ref{lem:surrogate}). Finally, we apply Theorem~\ref{thm:cardinal-goods} to these valuations and expand each allocated group into its constituent goods.

\subsection{Grouping the Original Goods}
Before constructing the auxiliary valuations, we partition the goods into groups. We then need only one value per group for each agent, and we can compare two collections of groups with a single query on their unions. Since the goods in each group must be allocated together, grouping can lower an agent's MMS. Fix a positive integer $k$. Following the compression approach of Lin, Liu, Tao, and Zhou~\cite{journals/corr/abs-2609-08687}, we control this loss by ensuring that every nonsingleton group is worth at most $\mu_i/k$ to each agent $i$.

If $m\le 4kn$, we take each good as a separate group. Otherwise, we fix an arbitrary order of the goods for all agents and adapt the recursive partitioning procedure of Lin, Liu, Tao, and Zhou~\cite{journals/corr/abs-2611-04753}. For each agent $i$, we start with $M$ as a single interval. At each step, we select a highest-valued nonsingleton interval $I$ and split it at a \emph{weighted median}, a good such that the goods strictly to its left and those strictly to its right in $I$ each have total value at most $v_i(I)/2$. To find such a good, we use their binary search, which compares prefixes of $I$ with their complementary suffixes and takes $O(\log |I|)$ comparisons. In case of a tie, the prefix and suffix are each worth $v_i(I)/2$, so either answer is valid. We keep the median as a singleton and treat each nonempty side as a separate interval. For each agent, we stop after $4kn$ splits or when all her intervals are singletons. We collect the cut points from all agents to form a single partition of $M$.

\begin{lemma}\label{lem:rank-groups}
For every positive integer $k$, the grouping procedure partitions $M$ into at most $\min\{m,8kn^2+1\}$ groups using $O(k^2n^3+kn^2\log m)$ bundle comparisons. When each group is treated as one good, every agent $i$ has MMS at least $(1-1/k)\mu_i$.
\end{lemma}
\begin{proof}
We first bound the number of groups and the number of comparisons needed to construct them. If $m\le 4kn$, we take each good as a separate group, which preserves every agent's MMS and requires no comparisons. Suppose that $m>4kn$. For each agent, we make at most $4kn$ splits, each introducing at most two cut points. As in the proof of Lemma~B.2 of Lin, Liu, Tao, and Zhou~\cite{journals/corr/abs-2611-04753}, we count the groups by collecting the cut points from all agents. There are at most $8kn^2$ distinct cut points, so they divide $M$ into at most $8kn^2+1$ groups. Since every group is nonempty, there are also at most $m$ groups. For a single agent, there are at most $8kn+1$ intervals at any step. Choosing a highest-valued nonsingleton interval therefore takes $O(kn)$ comparisons, and finding its weighted median takes $O(\log m)$ more by Lemma~B.1 of the same paper. Over at most $4kn$ splits for each of the $n$ agents, this gives $O(k^2n^3+kn^2\log m)$ comparisons.

We now bound the loss in MMS. Fix an agent $i$. Since the common partition includes all of her cut points, every nonsingleton group lies within one of her remaining nonsingleton intervals. If no such interval remains, all groups are singletons and her MMS is unchanged. Otherwise, let $x$ be the maximum value of her remaining nonsingleton intervals. Our first step is to show that $x\le\mu_i/k$. To do so, we construct an $n$-partition of $M$ in which every bundle is worth at least $kx$ to agent $i$.
Since a nonsingleton interval remains, the procedure made all $4kn$ splits for agent $i$. At each step, it split a highest-valued nonsingleton interval. Each interval on either side of the weighted median was worth at most half the value of the interval being split, so the maximum value among the nonsingleton intervals never increased. It follows that every interval selected for splitting was worth at least $x$.

Consider the binary tree formed by these $4kn$ intervals. The children of an interval are those of its side intervals that were subsequently split. We use the tree to find disjoint sets of goods, each worth at least $x/2$. From each leaf, take its entire interval, which is worth at least $x$. From each interval $I$ with exactly one child $J$, take $I\setminus J$. By the choice of the weighted median, $v_i(J)\le v_i(I)/2$, so $v_i(I\setminus J)\ge v_i(I)/2\ge x/2$. The sets obtained this way are disjoint because intervals in different branches are disjoint, and the set taken at a node with one child excludes that child's entire interval.
To form the desired partition, it suffices to have $2kn$ such sets. In a binary tree, the number of nodes with two children is one less than the number of leaves. Thus, at least $2kn$ of the $4kn$ nodes have at most one child, and each gives one of the disjoint sets above. Place $2k$ of these sets in each of $n$ bundles and assign the remaining goods arbitrarily. Every bundle is worth at least $2k\cdot(x/2)=kx$ to agent $i$. Hence, $\mu_i\ge kx$, and every nonsingleton group is worth at most $x\le\mu_i/k$ to her.

It remains to show that keeping these groups together reduces agent $i$'s MMS by at most $\mu_i/k$. Fix one of her MMS partitions and keep each singleton group in its original bundle. As in the proof of Lemma~\ref{lem:compress}, assign the nonsingleton groups one at a time to a bundle whose current value falls furthest short of its original value. Since each reassigned group is worth at most $\mu_i/k$, the same argument shows that each bundle loses at most $\mu_i/k$ relative to its original value. Every original bundle was worth at least $\mu_i$, so the resulting partition has minimum bundle value at least $(1-1/k)\mu_i$.
\end{proof}

\subsection{Constructing the Auxiliary Valuations}\label{subsec:surrogate}
We next construct an auxiliary additive valuation for each agent on the groups from Lemma~\ref{lem:rank-groups}. We describe the construction for an arbitrary set $M$ of $m$ goods. When we apply it to the groups, each group is treated as one good and $\mu_i$ denotes agent $i$'s MMS on the grouped instance.
Fix $\eta\in(0,1/16)$, and let $\mu(w)$ denote the MMS under a nonnegative additive valuation $w$. We say that $w$ is \emph{$\eta$-faithful} for agent $i$ if every bundle $S$ with $w(S)\ge(1-\eta)\mu(w)$ satisfies $v_i(S)\ge(1-8\eta)\mu_i$. Thus, applying Theorem~\ref{thm:cardinal-goods} to $\eta$-faithful auxiliary valuations at accuracy $\eta$ gives each agent the desired guarantee under her true valuation.

We construct the auxiliary valuations separately for each agent. Fix an agent $i$. Starting from $w(e)=0$ for every good $e$, we update $w$ over a sequence of rounds. In each round, we first estimate $\mu(w)$ using Woeginger's machine-covering scheme~\cite{journals/orl/Woeginger97}. Let $p$ be the minimum bundle value in the resulting partition, so that $(1-\eta)\mu(w)\le p\le\mu(w)$. This step uses only the known values $w(e)$. If fewer than $n$ goods have positive value under $w$, we take $p=0$, since $\mu(w)=0$.
We use $p$ to choose two bundles. First, among all bundles $S$ with $w(S)\ge p-3\eta$, we choose a bundle $B$ of minimum true value. As we show below, this search includes every bundle meeting the $(1-\eta)$-MMS target under $w$. Thus, $v_i(B)$ is a lower bound on the true value of every such bundle. We record the pair $(w,B)$ before updating $w$.
To guide the update, we choose a bundle $A$ of maximum true value among all bundles $S$ with $w(S)\le p+3\eta$. We show below that $v_i(A)\ge\mu_i$. If $v_i(B)<(1-8\eta)\mu_i$, then $v_i(A)-v_i(B)>8\eta\mu_i$, while the search thresholds give $w(A)-w(B)\le 6\eta$. This motivates the following update, which we perform in every round. Set $D=\ceil{m/\eta}$. We increase $w(e)$ by $1/D$ for each $e\in A\setminus B$ and decrease it by $1/D$ for each $e\in B\setminus A$, keeping the updated values in $[0,1]$. If the resulting total exceeds $2n$, we subtract $1/D$ from every coordinate and replace negative values by zero.

We implement both searches by a knapsack dynamic program on the integer scores $Dw(e)$, processing the goods one at a time. For $A$, we discard bundles $S$ with $w(S)>p+3\eta$ and keep a bundle of maximum true value at each remaining attainable score. After processing all goods, we choose the most valuable of these bundles. For $B$, we merge all scores corresponding to $w(S)\ge p-3\eta$ into one state and keep a bundle of minimum true value at each state. The bundle in the merged state after all goods have been processed gives $B$. Each choice between two candidates requires one bundle comparison. By additivity, adding the same unprocessed good to both preserves their order under $v_i$.
After $T=1+\floor{m\min\{m,2n\}/\eta^2}$ rounds, we compare the recorded bundles $B$, select one of highest true value, and return the valuation recorded with it. We show below that at least one recorded bundle is worth at least $(1-8\eta)\mu_i$ to agent $i$. Thus, the valuation returned by the procedure is $\eta$-faithful for $i$.

\begin{lemma}\label{lem:surrogate}
For every fixed $\eta\in(0,1/16)$, the procedure computes an $\eta$-faithful nonnegative additive valuation $w_i$ for each agent $i\in N$ in $\operatorname{poly}(n,m)$ time using $O(nm^3\min\{m,n\}/\eta^3)$ bundle comparisons in total.
\end{lemma}
\begin{proof}
Fix an agent $i$. We first use the estimate $p$ to relate the choices of $A$ and $B$ to agent $i$'s MMS. The auxiliary item values start at zero and remain in $\{0,1/D,\dots,1\}$. In each round, an item value increases by at most $1/D$. If the total exceeds $2n$, we subtract $1/D$ from every item value and replace negative values by zero. After this correction, each item value is at most what it was at the start of the round. Thus, $w(M)\le 2n$ at the start of every round, so $\mu(w)\le w(M)/n\le 2$. Combining this bound with $(1-\eta)\mu(w)\le p\le\mu(w)$ gives $\mu(w)-2\eta\le p\le\mu(w)$. Now fix an MMS partition under $v_i$. By the definition of $\mu(w)$, some bundle in this partition is worth at most $\mu(w)\le p+2\eta$ under $w$. This bundle is feasible in the search for $A$ and has true value at least $\mu_i$, so $v_i(A)\ge\mu_i$. For $B$, every bundle $S$ with $w(S)\ge(1-\eta)\mu(w)$ satisfies $w(S)\ge\mu(w)-2\eta\ge p-2\eta$ and is therefore feasible in the search for $B$. Since $B$ has minimum true value among the feasible bundles, $v_i(S)\ge v_i(B)$. Thus, the valuation recorded in any round with $v_i(B)\ge(1-8\eta)\mu_i$ is $\eta$-faithful for agent $i$.

To show that such a round exists, we compare $w$ with a fixed additive valuation $w^*$. If $\mu_i=0$, every round satisfies the required bound. Suppose that $\mu_i>0$, and set $w^*(e)=\min\{1,v_i(e)/\mu_i\}$ for each good $e$. Every bundle worth at least $\mu_i$ under $v_i$ is worth at least one under $w^*$. We also have $w^*(M)\le 2n-1$. To see this, suppose that $w^*(M)>2n-1$. Form $n-1$ bundles in turn, adding goods to each until its value under $w^*$ first exceeds one. Each of these bundles is worth at most two, since each good is worth at most one. The remaining goods form a last bundle worth strictly more than one as well. Since $v_i(S)\ge\mu_i w^*(S)$ for every bundle $S$, all $n$ bundles would be worth strictly more than $\mu_i$ under $v_i$, contradicting the definition of MMS.

We now show that every round with $v_i(B)<(1-8\eta)\mu_i$ decreases the squared distance $\|w-w^*\|_2^2$. In such a round, $w^*(A)\ge 1$ and $w^*(B)\le v_i(B)/\mu_i<1-8\eta$, while the search thresholds give $w(A)-w(B)\le 6\eta$. Thus, $w^*(A)-w^*(B)$ exceeds $w(A)-w(B)$ by more than $2\eta$. Before restricting the coordinates to $[0,1]$, the update adds $1/D$ on $A\setminus B$ and subtracts $1/D$ on $B\setminus A$. Expanding the squared distance and using the fact that at most $m$ coordinates change, we find that the decrease is at least
\begin{equation*}
    \frac{2}{D}\bigl[(w^*(A)-w^*(B))-(w(A)-w(B))\bigr]-\frac{m}{D^2}
    \ge \frac{4\eta}{D}-\frac{m}{D^2}
    \ge \frac{\eta^2}{m},
\end{equation*}
where the last inequality uses $m/\eta\le D\le 2m/\eta$. The remaining adjustments preserve this bound. Restricting each coordinate to $[0,1]$ cannot increase its distance from $w^*(e)$. If the resulting total exceeds $2n$, it exceeds $w^*(M)$ by more than one. Subtracting $1/D$ from every coordinate then decreases the squared distance by at least $2/D-m/D^2>0$, and replacing negative coordinates by zero cannot increase it.

It remains to show that $T$ rounds suffice. Initially, $\|w-w^*\|_2^2=\sum_{e\in M}w^*(e)^2\le\min\{m,2n\}$. If $v_i(B)<(1-8\eta)\mu_i$ in every round, the total decrease over $T$ rounds would exceed this initial squared distance. Thus, at least one recorded bundle $B$ satisfies $v_i(B)\ge(1-8\eta)\mu_i$. The procedure selects a recorded bundle of maximum true value, so the valuation recorded with it is $\eta$-faithful for agent $i$. Finally, since $p\le 2$, each knapsack search has $O(D)$ possible scores and uses $O(mD)=O(m^2/\eta)$ comparisons. For each agent, the procedure runs for $O(m\min\{m,n\}/\eta^2)$ rounds and uses at most $T-1$ further comparisons to select a recorded bundle. Over all agents, this gives $O(nm^3\min\{m,n\}/\eta^3)$ bundle comparisons. For fixed $\eta$, both the number of rounds and the time per round are polynomial in $n$ and $m$.
\end{proof}

Finally, we apply the algorithm in Section~\ref{sec:dichotomy} to the auxiliary valuations on the groups and transfer its guarantee to the original goods.

\begin{theorem}\label{thm:main}
For every fixed $\varepsilon\in(0,1)$ and all sufficiently large $n$, a $(1-\varepsilon)$-MMS allocation of $m$ goods can be computed in polynomial time under full valuations, and in $\operatorname{poly}(n,m)$ time using $O\bigl(n^2\log m/\varepsilon+n^8/\varepsilon^6\bigr)$ bundle comparisons.
\end{theorem}
\begin{proof}
The full-valuation statement follows from Theorem~\ref{thm:cardinal-goods}. For the comparison statement, set $k=\ceil{8/\varepsilon}$ and choose $\eta\in(\varepsilon/64,\varepsilon/32)$. By Lemma~\ref{lem:rank-groups}, we can partition $M$ into $m'\le\min\{m,8kn^2+1\}$ groups using $O(n^3/\varepsilon^2+n^2\log m/\varepsilon)$ bundle comparisons. Let $\mu_i'$ denote agent $i$'s MMS when each group is treated as one good. Then, $\mu_i'\ge(1-1/k)\mu_i$ for every $i\in N$.
We next apply Lemma~\ref{lem:surrogate} to the $m'$ groups to construct an $\eta$-faithful auxiliary valuation $w_i$ for each agent $i$. Each comparison between bundles of groups is a single query on the corresponding unions of original goods. Since $m'=O(n^2/\varepsilon)$ and $\eta=\Theta(\varepsilon)$, this step uses $O\bigl(n(m')^3\min\{m',n\}/\eta^3\bigr)=O(n^8/\varepsilon^6)$ bundle comparisons.
We now apply Theorem~\ref{thm:cardinal-goods} to $(w_1,\dots,w_n)$ at accuracy $\eta$. Each agent $i$ receives a bundle of groups $Y_i$ with $w_i(Y_i)\ge(1-\eta)\mu(w_i)$. Let $X_i$ be the union of the groups in $Y_i$. By $\eta$-faithfulness and the bound on $\mu_i'$, we have
\begin{equation*}
    v_i(X_i)
    \ge (1-8\eta)\mu_i'
    \ge (1-8\eta)(1-1/k)\mu_i
    \ge (1-\varepsilon)\mu_i
    \qquad\text{for every }i\in N,
\end{equation*}
where the last inequality follows from $8\eta+1/k\le\varepsilon$.
We note that only the grouping and the construction of the auxiliary valuations require comparison queries. The algorithm therefore uses $O(n^2\log m/\varepsilon+n^8/\varepsilon^6)$ bundle comparisons and runs in polynomial time for fixed $\varepsilon$.
\end{proof}

\section{Extension to Indivisible Chores}\label{app:chores}
In this section, we extend our polynomial-time MMS approximation schemes to indivisible chores under both full costs and bundle comparisons. 
For chores, each agent $i\in N$ has a nonnegative additive cost function $c_i\colon 2^M\to\mathbb R_{\ge0}$. We define her maximin share by $\mu_i=\min_{\bP\in\pi_n(M)}\max_{j\in[n]} \{c_i(P_j)\}$.
An allocation $\bX$ is $(1+\varepsilon)$-MMS if $c_i(X_i)\le(1+\varepsilon)\mu_i$ for every $i\in N$.

In Appendix~\ref{subsec:chores-cardinal}, we give an \emph{equal-cardinality reduction} from chores to goods. 
Specifically, we construct a goods instance and, for each agent, a partition into bundles of a common size. 
We then apply Theorem~\ref{thm:cardinal-goods} to this instance and convert the resulting allocation into a $(1+\varepsilon)$-MMS allocation of the original chores. 
In Appendix~\ref{subsec:chores-comparisons}, we implement the reduction using bundle comparisons.

\begin{theorem}\label{thm:chores}
For every fixed $\varepsilon\in(0,1)$ and all sufficiently large $n$, a $(1+\varepsilon)$-MMS allocation of $m$ chores can be computed in polynomial time under explicit costs, and in $\operatorname{poly}(n,m)$ time using $O(n^2\log m/\varepsilon+n^8/\varepsilon^6)$ bundle comparisons.
\end{theorem}

\subsection{Allocating Chores with Full Costs}\label{subsec:chores-cardinal}

When costs are given explicitly, we use the algorithm for goods from Theorem~\ref{thm:cardinal-goods} to turn the agents' individual partitions into a single allocation of chores. 
Fix $\delta\in(0,1/8)$. Suppose that, for each agent $i$, we are given a partition $\bP^i=(P_1^i,\dots,P_n^i)$ with $c_i(P_j^i)\le 1$ for every $j\in[n]$. Our goal is to find an allocation $\bX$ with $c_i(X_i)\le 1+3\delta$ for every $i\in N$.

To construct the goods instance, we modify each partition $\bP^i$ so that all bundles have a common size $K$ while increasing their costs only slightly. We then convert costs into goods values by assigning each item a value equal to two minus its cost. The minimum bundle value in agent $i$'s modified partition gives a lower bound on her MMS in the constructed instance. This bound is large enough that the goods algorithm must give each agent at least $K$ items. There are exactly $Kn$ items, so each agent receives exactly $K$. The cost of her bundle is therefore $2K$ minus its goods value, which gives the desired cost bound of $1+3\delta$.

\begin{lemma}\label{lem:chore-cardinal}
We can give each agent $i\in N$ a bundle $X_i$ with $c_i(X_i)\le 1+3\delta$ in polynomial time.
\end{lemma}
\begin{proof}
We first use the IDO reduction for chores~\cite{conf/sigecom/HuangL21} to obtain an instance with ordered items $e_1,\dots,e_m$. For each agent $i$, we sort her costs so that $c_i(e_1)\ge\dots\ge c_i(e_m)$ and apply the same permutation to her partition $\bP^i$. Since each of the $n$ bundles in $\bP^i$ costs at most one, her total cost is at most $n$. We apply Lemma~\ref{lem:compress} with $h=\ceil{2n/\delta}$ to obtain at most $2h$ compressed items and, for each agent, a partition in which every bundle costs at most $1+2n/h\le 1+\delta$. Let $d_i(e)$ denote agent $i$'s cost of compressed item $e$.

We next bound the number of compressed items in every bundle while increasing its cost by at most $\delta$. Set $L_0=\ceil{4/\delta}+2$, $K_0=\ceil{(1+\delta)L_0/\delta}$, and $K=K_0+L_0$. We modify each agent's partition separately. In every bundle with more than $K_0$ compressed items, we keep the $K_0$ costliest and set the rest aside. Each compressed item set aside costs at most $(1+\delta)/K_0$, since each of the $K_0$ retained in its bundle costs at least as much. We redistribute the compressed items set aside cyclically among the $n$ bundles. Each bundle receives at most $\ceil{2h/n}\le L_0$ of them, of total cost at most $\delta$. Thus, every bundle now contains at most $K$ compressed items and costs at most $1+2\delta$.

We then add zero-cost dummy items until there are exactly $Kn$ items in total, counting both compressed items and dummies. In each agent's partition, we distribute the dummies so that every bundle has exactly $K$ items. We define additive goods valuations by $v_i(e)=2-d_i(e)$, where $d_i(e)=0$ for a dummy. These values are nonnegative because each compressed item costs at most $1+2\delta<2$ and each dummy has cost zero. The use of dummy items and complementary values to enforce equal bundle sizes also appears in the chores-to-goods reduction for pairwise maximin share~\cite[Lemma~1]{journals/corr/abs-2609-10493}. Each bundle in agent $i$'s partition has value at least $2K-1-2\delta$, so her MMS under $v_i$ is at least $2K-1-2\delta$.

We now apply Theorem~\ref{thm:cardinal-goods} with accuracy $\delta/(2K-1-2\delta)$. For sufficiently large $n$, each agent $i$ receives a bundle $Y_i$ satisfying
\begin{equation*}
    v_i(Y_i)
    \ge \left(1-\frac{\delta}{2K-1-2\delta}\right)(2K-1-2\delta)
    = 2K-1-3\delta
    > 2(K-1).
\end{equation*}
The value of each compressed item is at most $2$, and that of each dummy is exactly $2$. Thus, every agent receives at least $K$ items. Since there are exactly $Kn$ items in total, every agent receives exactly $K$. It follows that $d_i(Y_i)=2K-v_i(Y_i)\le 1+3\delta$.

Removing the dummies and expanding the compressed items gives an allocation $\bX'$ of the ordered items with the same costs. To obtain an allocation $\bX$ of the original chores, we process the ordered items $e_j$ for $j=m,\dots,1$. When $e_j\in X'_i$, agent $i$ takes her least costly remaining chore. Only $m-j$ chores have been assigned, so at least one of her $m-j+1$ least costly chores remains. The chore she takes therefore costs at most $c_i(e_j)$. Summing over $e_j\in X'_i$ gives $c_i(X_i)\le\sum_{e\in X'_i}c_i(e)\le 1+3\delta$ for every $i\in N$. Finally, we note that all steps take polynomial time.
\end{proof}

\begin{proof}[Proof of Theorem~\ref{thm:chores} for explicit costs]
Choose $\delta\in(0,1/8)$ such that $(1+3\delta)(1+\delta)\le 1+\varepsilon$. If some agent has zero cost for every chore, assigning all chores to her gives every agent cost zero. We may therefore assume that $\mu_i>0$ for every $i\in N$.

For each agent $i$, we treat the chores as jobs with processing times $c_i(e)$ on $n$ identical machines. Assigning the jobs to the machines is equivalent to partitioning the chores into $n$ bundles, so the optimal makespan is $\mu_i$. The approximation scheme of Hochbaum and Shmoys~\cite{journals/jacm/HochbaumS87} gives a partition $\bP^i$ whose maximum bundle cost $c_i^{\max}=\max_{j\in[n]}\{c_i(P_j^i)\}$ satisfies $c_i^{\max}\le(1+\delta)\mu_i$.
Dividing agent $i$'s costs by $c_i^{\max}$ makes every bundle in her partition cost at most one to her. Lemma~\ref{lem:chore-cardinal} then gives an allocation $\bX$ in which every agent incurs scaled cost at most $1+3\delta$. Restoring the original costs, we obtain
\begin{equation*}
    c_i(X_i)
    \le (1+3\delta)c_i^{\max}
    \le (1+3\delta)(1+\delta)\mu_i
    \le (1+\varepsilon)\mu_i
    \qquad\text{for every }i\in N. \qedhere
\end{equation*}
\end{proof}

\subsection{Allocating Chores with Bundle Comparisons}\label{subsec:chores-comparisons}

Given an agent $i\in N$ and two bundles of chores $X,Y\subseteq M$, the query $\Comp_i(X,Y)$ returns a bundle of smaller cost under $c_i$, breaking ties arbitrarily on each query. 
For fixed $n$ and $\varepsilon>0$, Lin, Liu, Tao, and Zhou~\cite{journals/corr/abs-2609-08687} obtained a $(13/11+\varepsilon)$-MMS allocation using $O(\log m)$ comparisons. To obtain the $(1+\varepsilon)$-MMS guarantee of Appendix~\ref{subsec:chores-cardinal}, we follow Appendix~\ref{sec:comparisons} and use comparisons to group the chores and construct an auxiliary additive cost function for each agent.

We set $k=\ceil{8/\varepsilon}$ and choose $\eta\in(\varepsilon/64,\varepsilon/32)$ as in Appendix~\ref{sec:comparisons}. 
Then we group the chores using the procedure of Lemma~\ref{lem:rank-groups}. For each agent, we repeatedly select a highest-cost nonsingleton interval and split it at a weighted median with respect to her costs. Let $\mu_i'$ denote agent $i$'s MMS when each group is treated as one chore. 

On these groups, we adapt the construction of Lemma~\ref{lem:surrogate} separately for each agent $i$. In each round, let $\mu(w)$ denote the chore MMS under the current auxiliary costs $w$. We use the scheduling scheme of Hochbaum and Shmoys~\cite{journals/jacm/HochbaumS87} to find a partition whose maximum bundle cost $p$ satisfies $\mu(w)\le p\le(1+\eta)\mu(w)$. 
As in Section~\ref{subsec:surrogate}, we then use $p$ to select two bundles via knapsack dynamic programming over the discretized auxiliary costs: a bundle $A$ of maximum true cost subject to $w(A)\le p+3\eta$, and a bundle $B$ of minimum true cost subject to $w(B)\ge p-3\eta$. As we show below, every bundle $S$ meeting the target $w(S)\le(1+\eta)\mu(w)$ has true cost at most $c_i(A)$, whereas $B$ satisfies $c_i(B)\le\mu_i'$ and guides the update of $w$.
We therefore record $(w,A)$ before updating $w$ as in Lemma~\ref{lem:surrogate}. After the prescribed number of rounds with the groups as items, we return the cost function $w_i$ recorded with the least costly $A$.

\begin{proof}[Proof of the comparison statement of Theorem~\ref{thm:chores}]
We first bound the increase in MMS caused by grouping. Fix an agent $i$. For any nonsingleton group $G$, the argument using the binary tree in the proof of Lemma~\ref{lem:rank-groups}, applied to $c_i$, gives $2kn$ disjoint sets of chores, each of cost at least $c_i(G)/2$. Thus, $c_i(G)\le c_i(M)/(kn)\le\mu_i/k$. Now fix an MMS partition for agent $i$. 
The reassignment argument in~\cite[Lemma~4.5]{journals/corr/abs-2609-08687} then gives $\mu_i'\le(1+1/k)\mu_i$.

We next show what the two searches guarantee in each round. The auxiliary costs sum to at most $2n$, and each group costs at most one under $w$. If we assign the groups one at a time to a bundle of minimum current auxiliary cost, the chosen bundle costs at most two before each assignment and at most three afterward. Hence, $\mu(w)\le 3$. It follows that every bundle $S$ with $w(S)\le(1+\eta)\mu(w)$ satisfies $w(S)\le p+3\eta$. Since $A$ has maximum true cost among these search candidates, $c_i(S)\le c_i(A)$. For $B$, consider an MMS partition of the groups under $c_i$. By the definition of $\mu(w)$, some bundle in this partition has auxiliary cost at least $\mu(w)\ge p-3\eta$. This bundle is feasible in the search for $B$ and has true cost at most $\mu_i'$, so $c_i(B)\le\mu_i'$.

It remains to show that some round records a bundle $A$ with $c_i(A)\le(1+8\eta)\mu_i'$. If $\mu_i'=0$, every group costs zero to agent $i$, and the bound holds in every round. Otherwise, set $w^*(G)=c_i(G)/\mu_i'$ for each group $G$. Each coordinate lies in $[0,1]$, and their sum is at most $n$. In any round with $c_i(A)>(1+8\eta)\mu_i'$, the bound on $c_i(B)$ gives $w^*(A)-w^*(B)>8\eta$, while the search thresholds give $w(A)-w(B)\le 6\eta$. The squared-distance argument in the proof of Lemma~\ref{lem:surrogate} therefore shows that some round satisfies $c_i(A)\le(1+8\eta)\mu_i'$. The least costly recorded $A$ satisfies this bound as well.

Finally, we apply the algorithm of Appendix~\ref{subsec:chores-cardinal} to the auxiliary costs at accuracy $\eta$. For sufficiently large $n$, each agent $i$ receives a bundle of groups $Y_i$ with $w_i(Y_i)\le(1+\eta)\mu(w_i)$. Let $X_i$ be the union of the groups in $Y_i$. By the choice of $w_i$ and the bound on $\mu_i'$, we obtain
\begin{equation*}
    c_i(X_i)
    \le (1+8\eta)\mu_i'
    \le (1+8\eta)(1+1/k)\mu_i
    \le (1+\varepsilon)\mu_i
    \qquad\text{for every }i\in N,
\end{equation*}
where the last inequality follows from $8\eta\le\varepsilon/4$, $1/k\le\varepsilon/8$, and $\varepsilon<1$. Only the grouping and the construction of the auxiliary costs require bundle comparisons. Both steps have the same query bounds as in Appendix~\ref{sec:comparisons}, so the algorithm uses $O(n^2\log m/\varepsilon+n^8/\varepsilon^6)$ bundle comparisons and runs in polynomial time for fixed $\varepsilon$.
\end{proof}

\section{Dependence on the Accuracy}\label{app:accuracy}

In this section, we first bound the number of agents required by Theorem~\ref{thm:cardinal-goods} as a function of the target accuracy. 
Then we use this bound to derive the rates for goods and chores in Theorems~\ref{thm:main} and~\ref{thm:chores}, respectively. 
After compressing the items, we write $\varepsilon'=1-\beta$ for the target loss and take $\sigma=1-\varepsilon'/2$. We require $L\ge C\cdot(\varepsilon')^{-8}\cdot\log L$ for a sufficiently large absolute constant $C$ and sufficiently large $n$. In particular, $(\varepsilon')^{-1}\le L^{1/8}$.

We choose the losses in normalization and compression so that $\varepsilon'$ is a constant fraction of the original target loss. After compression, we have $O(n/\varepsilon')$ items, so $m\le n^2$ for sufficiently large $n$. Thus, when $\Delta\ge 2R$, we can apply Lemma~\ref{lem:slack-rounding}.
For $\Delta<2R$, we have $\kappa=\ceil{2/\varepsilon'}$. Since $L^4\le n/2$ for sufficiently large $n$, every ordinary agent has $\zeta_i\ge 1/2$. The proof of Lemma~\ref{lem:valuable-prefix} then gives
\begin{equation*}
    p(M_2)\le\left(1+\ceil{4/\varepsilon'}\right)\cdot L^4
    \qquad\text{and}\qquad
    v_i(M_2)\le 4p(M_2)+\sigma
    \qquad \forall i\in N_o.
\end{equation*}
Thus, $T=O(L^4/\varepsilon')<n$ for sufficiently large $n$, so $M_1$ is nonempty. Since $\Delta<1536L^2$, we also have $|N_r|\le 1536n/L^2\le n/(4L)$.

For Theorem~\ref{thm:common-prefix-allocation}, we use the parameter definitions of Subsection~\ref{sec:prefix-allocation} and choose
\begin{equation*}
    d_1=\frac{\varepsilon'}{4\left(\kappa+1\right)},\qquad
    \xi=\tau=\frac{\varepsilon'}{8},\qquad
    d_2=\frac{\varepsilon'}{8\left(\kappa+1/\tau\right)}.
\end{equation*}
These choices give $(\kappa+1)\cdot d_1=\varepsilon'/4$ and $\kappa\cdot d_1+\xi+\tau+d_2/\tau\le 5\varepsilon'/8$, as required for the ordinary and backup agents. We have $d_1=\Theta((\varepsilon')^2)$, $d_2=\Theta((\varepsilon')^2)$, and $q_{\max}=O((\varepsilon')^{-2})$.

\paragraph{Counting records.}
We use $\bar F=\binom{q_{\max}+\kappa}{\kappa}\le 2^{q_{\max}+\kappa}$, so $\log_2\bar F=O((\varepsilon')^{-2})$. Using $(\varepsilon')^{-1}\le L^{1/8}$ in the definitions of $G_{\max}$ and $H_{\max}$, we get $\log_2 G_{\max}=O((\varepsilon')^{-4}\cdot\log L)$ and $\log_2 H_{\max}=O((\varepsilon')^{-2}\cdot\log L)$. We also have $T/d_1=O(L^4/(\varepsilon')^3)$ and $\log_2 m\le 2L$. Substituting these bounds into the formula for $\gamma$ in Appendix~\ref{app:bundle-selection} gives $\log_2 \gamma=O\!\left((\varepsilon')^{-4}\cdot\log L\right)$.
In each local record, we recover the nonincreasing list of rounded values in an initial interval from the count at each level. We store at most $1+\floor{1/d_1}$ counts per interval, each at most $2\gamma\cdot(1+\floor{1/d_1})$. Across the initial intervals we store $O((\varepsilon')^{-4})$ counts, each using $O((\varepsilon')^{-4}\cdot\log L)$ bits. For the recorded part of $M_2$, we store $O(1/d_2)$ counts, each at most $T/\tau=O(L^4/(\varepsilon')^2)$. These counts recover both the length of the recorded part and its nonincreasing list of values. They use $O((\varepsilon')^{-2}\cdot\log L)$ bits in total. Hence, the logarithm of the number of local records is $O((\varepsilon')^{-8}\cdot\log L)$.
By the choice of $C$, we can bound the base-two logarithms of $G$, $\gamma$, and the number of local records by $L/8$. Thus, $|N_a|\ge(n-|N_r|)/2^{L/8}\ge \gamma$. We also have $\gamma\le n/(4L)$ and $G\cdot\sqrt n\le n/(4L)$ for sufficiently large $n$. When we form $N_s$, the initial $\gamma$ members of $N_a$ and all of $N_r$ number at most $n/(2L)$. We bring the size up to $\ceil{n/L}$ and then move in at most $G\cdot\sqrt n$ agents from small groups. Consequently, $n/L\le |N_s|\le 2n/L$.

\medskip

Since $\sqrt \gamma\le 2^{L/16}<\sqrt n/L$, the rounding loss in each group is smaller than its initial excess.
By the proof of Lemma~\ref{lem:exceptional}, every agent $i\in N_s\setminus N_b$ who is still below $\beta$ on $M_2$ satisfies
\begin{equation*}
    n-v_i(M_1)\le
    \frac{\left(\beta+1\right)\cdot |N_s\setminus N_b|}{\xi}+\frac{T}{\tau}
    =O\!\left(\frac{n}{\varepsilon'\cdot L}+\frac{L^4}{(\varepsilon')^2}\right).
\end{equation*}
Using $(\varepsilon')^{-1}\le L^{1/8}$, we bound this by $\varepsilon'\cdot n/4$ for sufficiently large $n$. In the bound on the value of agent $i$'s favorite remaining bundle in Lemma~\ref{lem:quota-assignment}, we lose at most $\kappa\cdot d_1<\varepsilon'/4$ from the partially included interval. For the remaining two error terms, we have $\phi_g\cdot n\ge\sqrt n$ and
\begin{equation*}
    \frac{\kappa\cdot\sqrt \gamma}{\sqrt n}+\frac{2\kappa}{L}
    =O\!\left(\frac{2^{L/16}}{\varepsilon'\cdot\sqrt n}
              +\frac{1}{\varepsilon'\cdot L}\right)
    \le\frac{\varepsilon'}{4}.
\end{equation*}
Thus, the best remaining bundle gives agent $i$ value at least $1-3\varepsilon'/4>\beta$.

\paragraph{Consequences for Goods and Chores.}
For goods, the reductions in the proof of Theorem~\ref{thm:main} allow us to take $\varepsilon'=\Theta(\varepsilon)$ under both full valuations and bundle comparisons. Thus, $L\ge C\varepsilon^{-8}\log L$ suffices for the $(1-\varepsilon)$-MMS guarantee, for a sufficiently large absolute constant $C$.
For chores, the equal-cardinality reduction determines the accuracy required from the algorithm for goods. In the full-cost proof of Theorem~\ref{thm:chores}, we can take $\delta=\Theta(\varepsilon)$. The common bundle size in Lemma~\ref{lem:chore-cardinal} is then $K=\Theta(\delta^{-2})$, so the reduction applies Theorem~\ref{thm:cardinal-goods} with accuracy $\delta/(2K-1-2\delta)=\Theta(\varepsilon^3)$. Under bundle comparisons, we run the full-cost algorithm at accuracy $\eta=\Theta(\varepsilon)$, so the required goods accuracy is again $\Theta(\varepsilon^3)$. Substituting this accuracy into the bound for goods shows that $L\ge C\varepsilon^{-24}\log L$ suffices for the $(1+\varepsilon)$-MMS guarantee in both models, after increasing $C$ if necessary.
Since $L=\ceil{\log_2 n}$, we may take $\varepsilon$ to be a sufficiently large constant times $(\log L/L)^{1/8}$ for goods and $(\log L/L)^{1/24}$ for chores. We therefore obtain
\begin{equation*}
    1-\alpha^*_n
    = O\!\left(\left(\frac{\log\log n}{\log n}\right)^{1/8}\right)
    \qquad\text{and}\qquad
    \rho^*_n-1
    = O\!\left(\left(\frac{\log\log n}{\log n}\right)^{1/24}\right).
\end{equation*}
The polynomial running times in both theorems apply to each fixed $\varepsilon$.

\addtocontents{toc}{\protect\setcounter{tocdepth}{2}}

\end{document}